\documentstyle[eqsecnum,aps]{revtex}
\begin{document}
\draft
\preprint{SU-GP-93/3-1;gr-qc/9303019}
\title{Loop Representations for 2+1 Gravity on a Torus}
\author{Donald M. Marolf}
\address{Physics Department, Syracuse University, Syracuse, New York 13244}
\date{\today}
\maketitle

\begin{abstract}
We study the loop representation of the quantum theory
for 2+1 dimensional general
relativity on a manifold, $M = {\cal T}^2 \times {\cal R}$, where
${\cal T}^2$ is the torus,
and compare it with the connection representation for this system.  In
particular, we look at the loop transform in the part of the phase
space where the holonomies are boosts and study its
kernel.  This kernel is dense in the
connection representation and the transform is not continuous
with respect to the natural topologies, even in its domain of
definition.  Nonetheless, loop representations
isomorphic to the connection representation corresponding to this
part of the phase space can still be
constructed if due care is taken.  We present this construction
but note that certain ambiguities
remain; in particular, functions of loops cannot be uniquely
associated with functions of connections.
\end{abstract}
\pacs{SU-GP-93/3-1,gr-qc/9303019}

\section{Introduction}

In recent years the idea of a non-perturbative approach to
quantum gravity has become increasingly popular.  This is largely due to the
fact that no satisfactory
perturbative theory of quantum gravity has been found.  Indeed,  perturbative
quantum gravity fails even by the usual standards of quantum field theory:
that is, no renormalization scheme has been found that consistently
makes the individual terms in the perturbation series finite.  Efforts
to extend quantum gravity using ideas such as introducing higher derivative
terms in the action, requiring supersymmetry, etc., have failed as well.

Looking beyond the problems faced by these theories, a completely satisfactory
perturbative formulation of a fundamental theory would have to do more than
give
a perturbation series whose individual terms are finite;  it should give
a series that actually converges.  Even perturbative string theory does not
meet this requirement \cite{strings}.  As a result, the possibility that
certain
theories for which no perturbative quantization exists can be quantized
using non-perturbative methods becomes attractive and the possibility that
gravity itself is such a theory becomes even more so.

For these reasons interest in non-perturbative methods has grown and
a number of non-perturbative approaches to quantizing gravity has
been discussed.  These include approaches that aim toward integrating
numerically path integrals associated with general relativity both
in a covariant field formulation \cite{Bryce} or in a Regge calculus
style formulation \cite{Regge}.  Alternatively, nonperturbative canonical
quantization has been considered using an ADM-based formalism to
construct constraint operators that act on functions of three-metrics
to select physical states \cite{Tril}.

A great simplification in the canonical approach is achieved by using a
different set of variables to define the theory.  Instead of considering
metrics to be fundamental, this formulation is based on
tetrads and self-dual connections \cite{SD1,SD2}.  This formulation
defines constraint operators that are polynomial in the basic phase space
coordinates so that these constraints
may be much easier to impose to select physical
states than the metric based constraints.
In this formulation, it is the connection and not the tetrad
that is considered to define the configurations space.  One
consequence of this change of viewpoint is that
gauge invariant quantities in this formulation
naturally involve holonomies around
loops and their traces -- the Wilson loop functionals -- just
as in other gauge theories.

In this way, loops enter naturally into this formulation of quantum
gravity.  It is even possible to define quantum states in terms of
functions of loops.  This set of loop functions provides a
representation of operators also based on loops and is therefore called
a ``loop representation" \cite{trans,Rov}.  Much effort has gone into
the study of loop representations ranging from the explicit
construction of an infinite dimensional space of solutions to the
constraints \cite{trans,Rov,Blen,Gam},
to the construction and
regularization of the type of observables that are more familiar
in general relativity and of states
that reproduce certain properties of flat three dimensional spaces
\cite{Smo,Rov}.

However, the program of quantizing gravity using a loop representation is far
from complete and it is important to gain
intuitive insight into the process as a whole by carrying out this program
on simplified models and other simple systems.  This allows us not only
to gain perspective from which we can evaluate the program as a whole
but also to take note of subtleties that may become important in trying to
carry out this program in the full theory.
In particular, loop representations have been constructed and analyzed for
linear systems such as
Maxwell fields \cite{Ash2}
(U(1) connections) and linearized gravity \cite{Smo2} as well as for the
genuinely nonlinear case of
2+1 dimensional gravity on ${\cal T}^2 \times {\cal R}$\cite{two,otwo,Ash}.
We note, however,
that the part of 2+1 gravity on a torus that was analyzed in
\cite{two,otwo,Ash} in terms of a loop representation describes only
connections whose holonomies are rotations and not those whose
holonomies are boosts.  As a result, all of these models are based in effect on
a compact gauge group in contrast to the full theory of 3+1 gravity.

One of the objects studied in these simple models is the
so-called ``loop transform."  This transform is a formal map from
functions of connections to functions of loops.  The investigation of the loop
transform in these simple models is important because the loop
transform has been frequently used, at least heuristically, in the
study of the loop representation in the 3+1 theory \cite{trans,Smo,Rov}.
The idea is that
it would send a state described by a function of connections to
the same state's description in terms of functions of loops.  One
might hope that this loop transform could be used to show that
suitable such loop representations are isomorphic to
suitable connection representations and therefore to further justify
the use of loop representations.
These expectations are borne out in simple models including
2+1 gravity on a torus in the ``sector" in which the holonomies are
rotations.  There has also been a general attempt to make the
loop transform and well defined using Gel'fand
spectral theory \cite{AI}.

Our purpose here will be to reopen the study of the loop representations
of 2+1 gravity on a torus by concentrating instead on the part of the
theory where the holonomies are boosts.  This is important from a physical
point of view, since it is this part of the phase space that actually
corresponds to spacetimes in which the toroidal sections are spacelike
surfaces \cite{Mess,LaM}.  It is also important from the point of loop
representations since this part of the theory is a simple model with
a non-compact gauge group.  We will see that this leads to a quite
different loop representation as well as a distinct change in the
properties of the loop transform.

This paper begins with a description of classical 2+1 general relativity
formulated in terms of a connection and a triad.  In the spirit of
canonical quantization, we
examinethe canonical structure of this theory and a description of
the reduced phase space.  In particular, we look at the ``sector"
of the phase space in which the holonomies are all
rotations and the one where the holonomies
are all boosts.  With this description in terms of
holonomies, we then introduce the loop-dependent phase space
functions that define the classical ``loop algebra."  This is the algebra
that leads to loop representations in the quantum theory.

Using this classical description, we can define a quantum theory
using reduced phase space quantization in both a connection
representation and a loop representation for both of the above sectors
of the phase space.  As mentioned before, the loop transform is a
representation isomorphism between connection and loop representations in the
sector in which the holonomies are rotations.  However, as we demonstrate in
Appendix \ref{uip}, the
naively defined loop representation in the
sector where the holonomies are rotations is identical to
the one defined in the sector where the
holonomies are boosts.  This result is disturbing since
the connection representations and even the classical solutions
corresponding to these sectors have quite different properties.

To understand this issue, we investigate the loop transform in
the sector where the holonomies are boosts.  As might be expected
from our dilemma, the loop transform is no longer a
representation isomorphism in this sector.  Instead, the loop
transform has a non-trivial kernel which turns out to be so
large that it is dense in the connection representation.  The loop
transform is also far from continuous as a map from the connection
representation to functions of loops.  The proof of these statements
represents a digression from the main discussion and is therefore relegated
to Appendix \ref{proof}.

Despite these problems, we then show that loop representations isomorphic
to the connection representation can in fact be constructed in this
sector, though this requires extra care.  The construction of such
loop representations is based on the choice of a dense subspace
of the connection representation that satisfies a short list of properties.
The actual construction of such subspaces again represents a tangent to
the main discussion and is presented in Appendix \ref{Vs}.  Since each of
these representations is isomorphic to the connection representation,
this construction is independent of the subspace chosen up to
isomorphism.  Nonetheless, certain ambiguities remain in the loop
representation and we discuss these as well.  We note that
many of these ambiguities can be removed by extending our
representation from functions of loops to functions of so-called
``generalized loops."  Finally, we close with a analogy between
the loop transform and the so-called "Mellin Transform" and a
short discussion of
the possibility of extrapolating our results to 2+1 gravity on
on other compact spacial slices and to the full theory in 3+1
dimensions.

\section{Summary of the Classical Formulation}
\label{CT}

We now begin with a brief review of the classical connection formulation
of 2+1 dimensional general relativity following \cite{Ash}.
This will include a review of the canonical description which we
will use in section \ref{QT} for canonical quantization of this system.
Our canonical discussion will
illustrate the construction of the reduced phase space and its
division into ``sectors" on which we will base our quantum theory.
This quantum theory will be described in two ways: in terms of a
connection representation and in terms of a ``loop representation."  This
loop representation
is defined as a representation of the so-called ``loop algebra" and therefore
we
we will also introduce the classical functions
that define this algebra.  We will see that these functions can be used
to label points in the two phase space sectors on which we will
concentrate.

Our classical description begins with the basic
variables of our theory; the co-triads, $^3e_a^I$, and
$SU(1,1)$ connections, $^3A_a^I$.  Instead of $SU(1,1)$, we could
use $SO(2,1)$ or its connected component as our gauge
group without significantly changing the discussion in sections
\ref{QT} - \ref{ip}.  \cite{LaM}  These variables are to be subject to
the dynamics specified by the action:

\begin{equation}
S(^{3}e_{a}^{I},\ ^{3}A_{b}^{I}) = \int_{M} {d^3x}
  \tilde{\eta}^{abc}
\ ^3{e}_{a}^{I}\ ^{3}F_{bcI}
\end{equation}
where $\tilde{\eta}^{abc}$ is the Levi-Civita density on
a three-dimensional manifold $M$ and

\begin{equation}
{}^3{F}_{abI} = 2 \partial_{[a}{}^{3}A_{b]I} +
\epsilon_{IJK}{}^{3}A_{a}^{J}{}^{3}A_{b}^{K}
\end{equation}
is the curvature of the connection, ${}^{3}A_{b}^{I}$.
Internal indices, I,J,..., are raised and lowered using the
2+1 Minkowski metric.
This action is analogous to the so-called ``self-dual action" in
3+1 gravity \cite{Bengston}.  As in that case, one can allow the triad to
become degenerate, although for nondegenerate triads the
theory is equivalent to 2+1 general relativity.

The dynamics are described more transparently through the classical
equations of motion:

\begin{equation}
{}^{3}{\cal D}_{[a}{}^3{e}_{b]I} = 0 \ \text{and} \
{}^3{F}_{ab}^{I} = 0
\end{equation}
Here, ${}^{3}{\cal D}$ is the gauge covariant derivative
operator determined by ${}^{3}A_{a}^{I}$, so that these equations
just say that the connection is compatible with the triad and than
the curvature of this connection is flat.  Note that, as defined,
${}^{3}{\cal D}$ acts only on internal indices.  We could
extend ${\cal D}$ to act on spacetime
indices in a number of ways, such as by requiring
compatibility with the triad, though this is not necessary.

Since we are interested in a canonical formulation of
this theory, we now  consider the Legendre transform of the
above action.  We assume that $M$ is of the form $\Sigma \times {\cal R}$
and take the space of pull-backs,
$A_{a}^{I}$, of the connection, ${}^{3}A_{a}^{I}$, to a
spacial slice, $\Sigma$, to be our the configuration space.
The momenta, $\tilde{E}_I^a$, are then duals of the pull-backs of the
co-triads: $\tilde{E}_I^a = \tilde{\eta}^{ab}e_{bI}$,
where $\tilde{\eta}^{ab}$ is the anti-symmetric Levi-Civita
density on $\Sigma$ and $e_{bI}$ is the pull-back of
${}^3e_{bI}$ to $\Sigma$.  The other variables are
Lagrange multipliers that enforce the constraints:

\begin{equation}
\label{constraints}
{\cal D}_{a}\tilde{E}_I^a = 0 \ \text{and} \
{F}_{ab}^{I} = 0
\end{equation}
where $F_{ab}^{I}$ is the curvature of $A_{a}^{I}$ and
${\cal D}$ is this connection's covariant derivative
operator.

The first constraint in \ref{constraints} generates internal $SU(1,1)$
transformations while the second generates
diffeomorphisms and the usual Hamiltonian
gauge transformation in 2+1 gravity.  This second constraint actually
generates an extension of these transformations to
degenerate triads.  Note that this second constraint depends only on
the connection and not on the conjugate momenta.  It follows that
any function of connections is invariant under transformations generated
by this constraint and that any function of connections that is invariant
under internal gauge transformations is a Dirac observable and
has vanishing Poisson Brackets
with all constraints.

We will use such gauge invariant functions of connections to
label the reduced configuration space of our system.  This will
allow us to describe the reduced phase space since
the reduced phase space will be the cotangent
bundle over the reduced configuration space.  From Eq. \ref{constraints},
we see that our reduced configuration space is the space
of $SU(1,1)$ equivalence classes of flat connections on
$\Sigma$.

A gauge equivalence class of connections on $\Sigma$ can
be characterized by its (equivalence class of)
holonomies, $U(\alpha)$, around all closed loops $\alpha$.
Holonomies have much less gauge freedom than connections and so
provide a step toward a reduced configuration space description.
Since the connection is flat, the holonomies around all
homotopically trivial loops are the identity
transformation and $U(\alpha)$ depends only on the
homotopy class, [$\alpha$], of $\alpha$.  This
set of holonomies forms a homomorphism from the homotopy group to
$SU(1,1)$ since $U([\beta] \circ [\gamma]) =
U([\beta]) U([\gamma])$.

The homotopy group of our surface, the torus, is
abelian with $\pi_1({\cal T}^2) = {\bf {\cal Z}} \oplus
{\bf {\cal Z}}$.  Since the set of holonomies forms
a homomorphism, it follows that for all closed loops
$\alpha$, $U([\alpha])$ is determined by the holonomies
around the two generators, $\alpha_1$ and $\alpha_2$, of
$\pi_1({\cal T}^2)$ and that $U([\alpha_1])$ and
$U([\alpha_2])$ commute.  Therefore, gauge equivalence
classes of flat connections on the torus are
labeled by pairs of commuting $SU(1,1)$ transformations.

We would like to find a simple parameterization of these holonomies.
Note that each holonomy $U[\alpha_b]$
is an $SU(1,1)$ transformation that leaves invariant
some internal 2+1 Lorentz vector $A_b^I$ and that an
infinitesimal transformation
that leaves this vector invariant is generated by the
element $A_b^I \tau_I$ in the Lie algebra of $SU(1,1)$.  Here, $\{\tau_I\}$
is basis for this Lie algebra that satisfies:

\begin{mathletters}
\begin{equation}
[\tau_0,\tau_1] = -i \case{1}{2} \tau_2
\end{equation}
\begin{equation}
[\tau_2,\tau_0] = -i \case{1}{2} \tau_1
\end{equation}
\begin{equation}
[\tau_1,\tau_2] = i \case{1}{2} \tau_0
\end{equation}
\end{mathletters}
Some finite transformations that leave this vector invariant are given by
\begin{equation}
\label{exp}
g = \exp{(\lambda  A_b^I  \tau_I)}.
\end{equation}
for $\lambda \in {\cal R}$.

Unfortunately, not all elements of $SU(1,1)$ can be written
as the exponential of some element of the Lie Algebra \cite{LaM}.
If a transformation preserves a
timelike vector, then it can be written in the form of Eq. \ref{exp}
for appropriate $A_b^I$
but this is not necessarily true if the transformation preserves a
spacelike or null vector.  If the transformation preserves a
spacelike or null vector, then this
transformation can either be written in the form of Eq. \ref{exp} or it
can be written as an element of the form given in Eq. \ref{exp}
multiplied by the group element $R(2 \pi)$.  $R(2 \pi)$ is the
group element that when acting on spinors induces a rotation
through the angle $2 \pi$ about any spacelike axis.  In the
two-dimensional representation  of $SU(1,1)$ $R (2 \pi)$ is
represented by the matrix $-\openone$.  Because of this, and because
$R(2 \pi)$ commutes with all $g \in SU(1,1)$, we will label group
elements which when acting on vectors preserve some null or spacelike
vector by the pair

\begin{equation}
g_b = (A_b^I, \epsilon_b) = \exp{(A_b^I \tau_I)}
\ [R(2 \pi)]^{({{\epsilon_b
-1} \over {2}})}
\end{equation}
All elements of $SU(1,1)$ for which $A_b^I$ is spacelike
or null are uniquely labeled by this
prescription.  Group elements that preserve a timelike vector can
also be labelled in this way but do not require the extra index,
$\epsilon_b$.

We would now like to see what restrictions we place on our
parameters when we require that the holonomies commute.
Two $SU(1,1)$ transformations commute only if they both preserve the same
2+1 dimensional vector so $A_1^I$ and $A_2^I$ must both
be proportional to the same internal vector, $t^I$:
\begin{equation}
A_b^I = 2a_b t^I
\end{equation}
The $\epsilon_b$ labels that arise when $t^I$ is spacelike or null
are unrestricted by this requirement.

The next step is to translate the gauge transformations of the holonomies
into transformations on these internal vectors.  The
sets of holonomies fall into equivalence classes
under gauge transformations which act on the set of holonomies by
conjugation with an element of $SU(1,1)$:

\begin{equation}
\{U[\alpha]\} \rightarrow \{ g U[\alpha] g^{-1} \}
\end{equation}
Since $R(2 \pi)$ commutes with every $SU(1,1)$ transformation, this
will have no effect on the labels $\epsilon_b$ that may be required.
On the other hand, conjugation of $\exp{(A_b^I \tau_I)}$
by the group element, $g$, yields the element, $\exp{(\tilde{A}_b^I \tau_I)}$,
where $\tilde{A}_b^I$ is obtained from $A_b^I$ by the action of the group
element $g$ on internal vectors
since our internal vectors label elements
of the Lie algebra of $SU(1,1)$.  Thus, equivalence classes of pairs of
holonomies are labeled by equivalence classes of pairs of
parallel internal vectors under $SU(1,1)$
transformations.

To classify these equivalence classes completely note that
for timelike or spacelike $t$ we
can choose $a_b$ so that the norm of $t$ is $\pm 1$.  Also note that for
$t$ null or timelike we can always choose $a_b$ to have
appropriate signs so that $t$ is future directed.
Now, all normalized future directed timelike $t$'s are related by
$SU(1,1)$ transformations, as are all normalized spacelike $t$'s and all
future directed null $t$'s.   Following
Ref. \cite{Ash} we will refer to the regions of reduced configuration
space corresponding to these three equivalence classes as the
timelike, spacelike, and null ``sectors" of
the theory and proceed to consider them separately.  (For
further details, see Ref. \cite{LaM}).

In each of these sectors, we would like to find an explicit
parameterization of the configuration space.  Suppose that, in each
sector, we fix a representative internal vector $t^I$ subject to the
above conditions.
Note that in the spacelike sector the holonomies are boosts so that
each $a_b$ in ${\bf {\cal
R}}$ labels a distinct $SU(1,1)$ transformation, $(2a_b t^I, \epsilon_b)$.
However, given any spacelike vector, $t^I$, there is an
$SU(1,1)$ transformation that
can reverse its direction.  Thus we can change
$(2a_bt^I, \epsilon_b)$ into $(-2a_bt^I, \epsilon_b)$
by a gauge transformation and, since we defined our parameterization by
regarding $t^I$ as
fixed, we should identify $\{a_b\}$ and $\{-a_b\}$ as gauge related.
No other values of these parameters give rise to gauge related holonomies so
the reduced configuration space in the spacelike sector is labeled
by $\{ a_b \} \in {\cal R}^2/{\cal Z}^2$ and $\{ \epsilon_b \}
\in \{ -1,1\} $.

To find the reduced configuration space in the null
sector, we note that the $SU(1,1)$ transformations
$\exp{(2a_b t^I \tau_I)}$ are distinct
for $\{ a_b \} \in {\cal R}^2$.  However, we must identify them
in gauge equivalence classes under $SU(1,1)$ transformations.
$SU(1,1)$ transformations can scale any null
vector $2a_b t^I$ by any
positive real number but leave the labels $\epsilon_b$ unaffected.  If we
ignore the equivalence class labeled by $\{ a_b \} = 0$ then since
$t^I$ is fixed each
equivalence class is uniquely labeled by an angular coordinate.  The
reduced configuration space for the null sector is
thus $S^1 \times ({\cal Z}_2)^2$
together with
four additional points representing the equivalence classes corresponding
to the pairs of holonomies $(\pm \openone, \pm \openone)$.

Finally, in the timelike sector,
the holonomies are
rotations about some axis.  Distinct rotations in $SU(1,1)$ are
labeled by angles defined modulo $4 \pi $.  For a rotation, $\exp{(\theta
\tau_0)}$
around the 0-axis, this angle is just the coefficient, $\theta$.  If we fix
$t^I$ in the timelike sector to point along the 0-axis, it
follows that our labels $a_1$ and $a_2$ in the configurations space are both
periodic with period $2 \pi$ and that the
configuration space is $a_b \in {\cal T}^2$.
Note that all of these sectors overlap at the zero
connection ($a_b = 0$).  The zero connection represents a
special point in phase space, which will be discussed in \cite{LaM}.
This represents a technical point that we ignore here because it does not
substantially alter the discussion.

We are interested only in the spacelike and timelike sectors
and, following \cite{Ash}, we take the corresponding
reduced phase spaces to be
the cotangent bundles over ${\bf {\cal R}}^2$ and ${\cal
T}^2$ respectively.  This means that we include the zero
connection in both sectors and deal with the ${\bf {\cal
Z}}_2$ symmetry of the spacelike sector separately in
order to give the reduced phase space the structure of a
cotangent bundle over a manifold without boundary\footnote{For a
proper treatment of the reduced phase space, see \cite{LaM}.}.  This
ensures that the resulting quantum treatment of the
spacelike sector is the one that would result from Dirac
quantization.  Because we are treating the sectors as
disconnected and the four spacelike sectors (labeled by the
$\epsilon_b$) are similar, we will consider only the spacelike
sector for which the discrete indices take the values
$\epsilon_1 = +1 = \epsilon_2$.  This completes
the characterization of the reduced
configuration space that we have been working toward.

One final element of the classical theory that we will need for our quantum
description is the set of functions that make up the classical ``loop
algebra" \cite{Smo,Ash}.
The loop algebra is defined by the Poisson algebra of the
functions:

\begin{mathletters}
\label{connectionts}
\begin{equation}
\label{connectiont0}
{\cal T}^{0}[\alpha ](A) = \case{1}{2} Tr\,U(\alpha)
\end{equation}
\begin{equation}
\label{connectiont1}
{\cal T}^{1}[\alpha ](A, \tilde{E}) = \oint_\alpha {dS^a
Tr\,E_{a}U(\alpha)}
\end{equation}
\end{mathletters}
for all closed loops, $\alpha$, where $E_a = \overline{\eta}_{ab}
\tilde{E}^b$ ($\overline{\eta}$ is
the anti-symmetric density on $\Sigma$ of weight -1) and the
2-dimensional representation of $SU(1,1)$ is used to take
the trace.  It can be shown on general grounds that
these functions are overcomplete almost everywhere on the phase space
of connections identified in gauge equivalence classes under internal
gauge transformations and so can
be used to label points in this
phase space up to a small ambiguity \cite{Lew}.

Although these functions take arbitrary closed loops as one of their
arguments, we are considering only flat
connections and these functions therefore
depend only on homotopy classes of loops.  Since any
homotopy class $[\alpha]$ can be decomposed in terms of the
generators, $[\alpha] = n_1[\alpha_1] + n_2[\alpha_2]$,
it will be simplest to regard the ${\cal T}^0$'s and ${\cal T}^1$'s
as functions of pairs of integers, ${\bf n} = (n_1,n_2)$.  Eq.
\ref{connectionts}
shows that the ${\cal T}^0$'s and ${\cal T}^1$'s are all
invariant under reversal of the loops so that
distinct ${\cal T}^0$'s and ${\cal T}^1$'s are actually
labelled by ${\bf n} \in {\bf {\cal Z}}^2/{\bf
{\cal Z}}_2$.

Using this notation, we can write the Poisson algebra of
the ${\cal T}^0$'s and ${\cal T}^1$'s as

\begin{mathletters}
\label{comms}
\begin{equation}
\label{comm1}
\{{\cal T}^0({\bf k}), {\cal T}^0({\bf m})\} = 0
\end{equation}

\begin{equation}
\label{comm2}
\{{\cal T}^0
({\bf k}),
 {\cal T}^1
({\bf m})\} = i {\bf k} \times {\bf m} \
 ({\cal
T}^0({\bf k}
 + {\bf m}) - {\cal T}
^0({\bf k} - {\bf m}))
\end{equation}

\begin{equation}
\label{comm3}
\{{\cal T}^1({\bf k}), {\cal T}^1({\bf m})\} = i {\bf k} \times {\bf m}
\ ({\cal
T}^1({\bf k} + {\bf m}) - {\cal T}^1({\bf k} - {\bf m}))
\end{equation}
\end{mathletters}
where ${\bf k \times {\bf p}} = k_1 p^2 -
k_2 p^1$.  The ${\cal T}^0$'s and ${\cal T}^1$'s also obey the relations
\begin{mathletters}
\label{allalgrels}
\begin{equation}
\label{algrel1}
{\cal T}^0({\bf k}){\cal T}^0({\bf m}) = \case{1}{2}
({\cal T}^0({\bf k} + {\bf m}) + {\cal T}^0({\bf k} - {\bf
m}))  \end{equation}

\begin{equation}
\label{algrel2}
{\cal T}^0({\bf k}){\cal T}^1({\bf m}) + {\cal T}^0({\bf
m}){\cal T}^1({\bf k}) = \case{1}{2} ({\cal T}^1({\bf k} +
{\bf m}) + {\cal T}^1({\bf k} - {\bf m}))
\end{equation}
\end{mathletters}

Note that the ${\cal T}^0$'s are functions of connections only and that
that are invariant under internal gauge transformations.  They are therefore
invariant under all gauge transformations and describe functions on the
reduced configuration space.  The ${\cal T}^1$'s are invariant under
internal gauge transformations but involve the momenta as well as the
connection so that their Poisson Brackets with the
the $F_{ab}^I = 0$ constraint
must be checked.  However, explicit calculation shows that
these Poisson Brackets vanish.  It
follows that the ${\cal T}^1$'s are also functions on the reduced phase space.
Both ${\cal T}^0$'s and ${\cal T}^1$'s
can therefore be written in terms of the phase space coordinates described
above:

\begin{mathletters}
\label{timedefs}
\begin{equation}
\label{t0time}
{\cal T}^0({\bf k}) = \cos{({\bf k \cdot a})}
\end{equation}
\begin{equation}
\label{t1time}
{\cal T}^1({\bf k}) =  \bigl( \sin{({\bf k \cdot a})} \bigr) \ {\bf k
\times {\bf p}}
\end{equation}
\end{mathletters}
in the timelike sector and by

\begin{mathletters}
\label{spacedefs}
\begin{equation}
\label{t0space}
{\cal T}^0({\bf k}) = \cosh{({\bf k \cdot a})}
\end{equation}
\begin{equation}
\label{t1space}
{\cal T}^1({\bf k}) = \bigl(  \sinh{({\bf k \cdot a})} \bigr) \ {\bf k
\times {\bf p}}
\end{equation}
\end{mathletters}
in the spacelike sector, where the bold-faced letters represent two component
vectors and $p^b$ is the momentum
conjugate to $a_b$.

Note that because the ${\cal T}^0$'s and ${\cal
T}^1$'s are invariant under ${\bf a}\ \rightarrow \ -{\bf
a}$, they are two-fold degenerate in both sectors.  This
will reintroduce the desired ${\bf {\cal Z}}_2$ symmetry
in the spacelike sector and lead to a split in the quantum theory
of the connection
representation in the timelike
sector into two irreducible representations of the ${\cal
T}^0$, ${\cal T}^1$ algebra.  Other
than this, we see that the ${\cal T}^0$'s and ${\cal T}^1$'s are
indeed an overcomplete set of functions on the reduced
phase space except at the
zero connection in the spacelike sector and at $a_b \in
\{0,\pi\}$ in the timelike sector.

This completes our description of the classical reduced phase space.
We see that it can be considered to have several ``sectors," two of
which are the ``timelike" sector, in which the fundamental holonomies are
two commuting rotations and the ``spacelike" sector in which the
fundamental holonomies are two commuting boosts.  The reduced configuration
space is a torus in the timelike sector and a plane in the spacelike
sector.  In each sector, we introduced additional functions on the phase
space, ${\cal T}^0$,
${\cal T}^1$, for each homotopy class that define the classical loop
algebra and can be used to
label points on the reduced phase space up to a small
number of degeneracies.

\section{The Quantum Theory}
\label{QT}

We
could build a quantum theory using either reduced phase
space quantization or Dirac constraint quantization.  In this case,
(with proper choices of factor ordering) we would find equivalent
results. (Compare our results with \cite{Ash}.)  Because the above
constraints are linear in or independent of momenta,
this is what would be expected (see, eg.
\cite{Kuchar}), up to questions about the gauge orbits.
We will choose reduced phase space quantization, which will allow
us to use the classical discussion of the last section to construct
a connection representation of
the quantum theory
in a straightforward manner.

A connection representation is defined as a representation of some
algebra on a Hilbert
space of functions of connections.  Since our equivalence
classes of connections are labeled in each sector by $\{ a_b \}$ we will
define our connection representation to consist of
functions of $a_b$ that are square integrable
with respect to some measure.  This measure must be determined by some
additional requirements.  If we wish to construct a representation of
$a_b$ and $p^b$ then it is reasonable to
to require $a_b$ and
$p^b$ to be Hermitian from which it follows that the measure must be
$d\mu = d^2a$.  If we take the ${\cal T}^0$'s and ${\cal
T}^1$'s to be given by

\begin{mathletters}
\label{hb1}
\begin{equation}
{\cal T}^0({\bf k}) = \cos{({\bf k \cdot a})}
\end{equation}
\begin{equation}
{\cal T}^1({\bf k}) =  \bigl( \sin{({\bf k \cdot a})} \bigr) \ {\bf k}
\times \bigl( -i {{\partial} \over {\partial {\bf a}}} \bigr)
\end{equation}
\end{mathletters}
in the timelike sector and by

\begin{mathletters}
\label{hb2}
\begin{equation}
{\cal T}^0({\bf k}) = \cosh{({\bf k \cdot a})}
\end{equation}
\begin{equation}
{\cal T}^1({\bf k}) = \bigl(  \sinh{({\bf k \cdot a})} \bigr) \ {\bf k}
\times \bigl( -i {{\partial} \over {\partial {\bf a}}} \bigr)
\end{equation}
\end{mathletters}
in the spacelike sector, then the ${\cal T}^0$,
${\cal T}^1$'s are also Hermitian
with respect to this inner product\footnote{In equations \ref{hb1} and
\ref{hb2} we have chosen units such
that $\hbar =1$ and we will continue to chose such units throughout our
discussion.}.  With this ordering,
the ${\cal T}^0$'s and ${\cal T}^1$'s again satisfy
Eq. \ref{allalgrels} and their commutation relations are
given by $i$ times the Poisson brackets ordered as in Eq.
\ref{comms}.  Thus, the Hilbert space of these connection representations
also carries a representation of the quantum loop algebra with
Hermitian ${\cal T}^0$'s and ${\cal T}^1$'s.  In fact, this same
Hilbert space would arise if we were to define the space as a connection
representation of the loop algebra (see Appendix \ref{Tip}).

Note that the ${\cal T}^0$, ${\cal T}^1$ algebra preserves the various
sectors.  Each sector must therefore provide a representation of this
algebra.  Because the ${\cal T}^0$'s and ${\cal T}^1$'s can be used
to construct the action of ${\bf a}$
and ${\bf p}$ on a dense set of states in the spacelike sector
of the connection
representation, no closed linear subspace of the carrier space
of the connection representation in the spacelike sector is left invariant
by the action of the  ${\cal T}^0$'s and ${\cal T}^1$'s.  In other words,
the spacelike connection representation forms
a topologically irreducible representation of the loop algebra.
However, in the timelike
sector this construction is possible only up to
an ambiguity between ${\bf a}$ and $-{\bf a}$.  As a result,
the timelike sector
contains two topologically irreducible
representations of the loop algebra.  One
of these is carried by the set of functions that are
symmetric under ${\bf a} \ \rightarrow \ -{\bf a}$ and the
other is carried by the set of functions that are antisymmetric under
this operation.  We will henceforth refer to these sets of functions
as the symmetric and
antisymmetric timelike sectors.

This completes our discussion of the connection representations.
Indeed, our goal is to examine the
so-called loop representation so we now proceed in this direction.
We begin with the definition of a loop representation: a loop
representation is a representation of the ${\cal T}^0$, ${\cal T}^1$
algebra on a Hilbert space which is obtained by completion
in the topology defined by some inner product of some linear
space whose elements are suitable
functions of loops in a manifold
$\Sigma$\footnote{This is a consequence of the definition given in
\cite{Ash}.}.  Usually, this linear space is taken to be
the set of functions with support
on a finite number of loops.

``Suitable" functions of loops
are to be constant on equivalence classes of loops
defined as follows:
two closed loops
$\alpha$ and $\beta$ are
equivalent if and only if $TrU[{\alpha}](A) = TrU[{\beta}](A)$ for
all connections $A$ in whatever class of connections we
are considering \cite{Ash}.  These
equivalence classes are called equitopy classes.  Since we
are concerned here only with flat connections on ${\cal
T}^2$ and since for this class of connections $U({\alpha})$
depends only on the homotopy class of $\alpha$, our
representation will be in terms of functions of homotopy
classes of loops in ${\cal T}^2$.  In addition, inverting
a loop ($\alpha \rightarrow - \alpha$) does not change
the trace of the holonomy around it.  Thus, we consider functions of
loops that are invariant under inversion of loops.  Such functions are
completely characterized by a function $\psi:{\cal R}^2/{\cal Z}_2
\rightarrow {\cal C}$.

These functions of loops\footnote{In the rest of this section and the next,
two equitopy classes of
loops will be referred to
simply as ``loops."} carry a representation of the ${\cal
T}^0({\bf k})$, ${\cal T}^1({\bf k})$'s, which are taken to be the
fundamental operators in this ``loop representation".  The action of
these operators on functions of loops is given by:

\begin{mathletters}
\label{loopts}
\begin{equation}
\label{t0}
({\cal T}^0({\bf k}) \psi)({\bf n}) = (\psi({\bf n} +
{\bf k}) + \psi({\bf n} - {\bf k}))/2
\end{equation}
\begin{equation}
\label{t1}
({\cal T}^1({\bf k}) \psi)({\bf n}) = i {\bf k}
\times {\bf n} \ (\psi({\bf n} + {\bf k}) - \psi({\bf n} -
{\bf k}))
\end{equation}
\end{mathletters}
With these definitions, the ${\cal T}^0$'s and
${\cal T}^1$ again satisfy commutation
relations given by $i$ times the Poisson brackets in
Eq. \ref{comms} and the algebraic relations given in Eq. \ref{allalgrels}.

The loop representation can be
formally related to the connection
representation via the ``loop transform" (see Ref. \cite{trans}) which
defines a function of loops from a function of connections.  The loop
transform has the general form:

\begin{equation}
\psi (\alpha) \equiv \int d\mu (A)\ \psi (A) {\cal T}^0[\alpha](A)
\end{equation}
which, in our case, is just

\begin{equation}
\psi ({\bf n}) = \int {d^2a \ \psi ({\bf a}) \ {\cal
T}^0({\bf n})}
\end{equation}
Formal arguments indicate in general that the loop transform
maps ${\cal T}^0$'s and ${\cal T}^1$'s in the connection
representation to ${\cal T}^0$'s and ${\cal T}^1$'s in the
loop representation \cite{trans}.
These arguments are rigorous in the timelike and
spacelike sectors when these operators act on
functions $\psi({\bf a})$ of connections on which
the loop transform is defined or functions $\hat{\psi}({\bf n})$ of loops
in the image of the loop transform.

While the loop transform can be used to relate the connection
and loop representations, this relationship takes different forms in different
sectors.  For example, the ${\cal T}^0({\bf k})$'s are symmetric under
inversion of loops so that the loop
transform annihilates any function of
connections in the antisymmetric timelike sector.
However, in the symmetric timelike sector, the loop transform
is just the Fourier transform so that if we impose the
$l^2$ inner product on the functions of equitopy classes of loops, the loop
transform is a representation isomorphism.  In fact, the
$l^2$ inner product is the only inner product
on the functions of loops for which ${\cal T}^0$,
${\cal T}^1$'s are Hermitian and the representation is
topologically irreducible if we suitably restrict what we mean by a
loop representation.  In particular we show in Appendix \ref{uip}
that this follows if we require
the zero function of loops to represent the
zero state and the representation to contain at least one function
with support on a single loop.  In this sense,
the most natural inner product to impose on a loop representation gives
a representation which is isomorphic to the connection representation in
the symmetric timelike sector.

We would like to ensure that the representations of the ${\cal T}^0$ and
${\cal T}^1$ operators in a loop representation capture enough of the
properties of the classical  ${\cal T}^0$ and
${\cal T}^1$ functions that we would have some hope that a quantum theory
based on this representation could produce the correct classical limit.
In particular, classically the  ${\cal T}^0$ and
${\cal T}^1$ functions are subject to a number of relations that include
both linear relations and inequalities (see \cite{Loll}).  Quantum versions
of these relations appear in the connection representation.  Yet, it
is difficult to see how we could ensure
that anything like this set of relations
appears in a loop representation, unless we know that the loop
representation is isomorphic to a connection representation.
For this purpose and others (see \cite{Smo,trans}) it would be nice if
the loop transform was always a representation
isomorphism, at least when
the ${\cal T}^0$'s and ${\cal T}^1$'s form an almost
complete set of operators as they do in the spacelike and symmetric
timelike sectors.  In such cases, we could then do away with the connection
representation entirely and use only the loop
representation.  Let us suppose for a moment that this is
true.

Notice that the loop representation is exactly the same
in both the spacelike and timelike sectors.  Both sectors
lead to the representation (\ref{loopts}) of the loop algebra as well
as to the same definitions of equitopy classes of
loops and therefore to the same class of functions of
loops.  As indicated above, the $l^2$ inner product is
the only inner product in this
representation compatible with the assertion that ${\cal
T}^0$'s and ${\cal T}^1$'s are Hermitian, as they are in both
the spacelike and timelike connection representations, and with the
assertions that the representation is topologically irreducible, contains
a function with support on a single loops, and is such that the zero function
of loops represents the zero state.  Unless there is additional input,
there is no remaining freedom to construct a loop representation.
If, then, the loop transform is a
representation isomorphism in both sectors, all of the above
requirements (except perhaps the existence of a function with
support on a single loop) would be satisfied in both sectors and we would
expect the two sectors to be isomorphic\footnote{Thanks to Jorma
Louko for pointing this out.}.

This result is disturbing.  Not only do the
two sectors have extremely different properties in terms
of classical spacetimes (see \cite{LaM}), but the
quantum theories in the connection representation
are distinct as well.  To see this, note that in
both representations, ${\cal T}^0({\bf k})$ are
multiplication operators.  However, in the timelike sector they
are represented by multiplication by functions of absolute
value equal to or less than one whereas in the spacelike
sector they are represented by multiplication by functions
of absolute value equal to or greater than one.  Thus the
spectrum $\sigma ({\cal
T}^0({\bf k}))$ of ${\cal T}^0({\bf k})$ is bounded in the timelike sector
($\sigma ({\cal T}^0({\bf k})) = [-1,1])$, but not in the
spacelike sector ($\sigma ({\cal T}^0({\bf k})) = [1,\infty)$ for
the $\epsilon_1 = +1 = \epsilon_2$ spacelike sector).
These representations are certainly not isomorphic.  We
therefore expect trouble with the loop transform in the
spacelike sector.

\section{the loop transform in the spacelike sector}
\label{LT}

We saw in the preceding section that the loop transform
has different properties in different
sectors of the theory.  For example, it annihilates the antisymmetric
timelike sector but is a representation isomorphism in the symmetric
timelike sector.  We also took a brief look at a dilemma that could
arise if the loop transform is a representation isomorphism in the
spacelike sector.  Therefore, we would now like to examine the
loop transform more carefully in this sector.  From this,
we should learn
about the loop transform in general and about the resolution of our
dilemma in particular.

For a careful study, we will need the explicit form of the
loop transform, ${\bf L}$, in the spacelike sector:

\begin{equation}
({\bf L}(\psi))({\bf n}) = \int{d^2a \ \cosh{({\bf
n \cdot a})}} \psi ({\bf a})
\end{equation}
Note that ${\bf L}$ is just the (two-sided)
Laplace transform in this sector due to the symmetry of
the functions of connections.  Explicitly, ${\bf L}$
is the two-sided Laplace transform in each coordinate, $(a_1,
a_2)$:

\begin{equation}
({\bf L} \psi)({\bf n}) = \int_{-\infty}^{\infty} da_1
\int_{-\infty}^{\infty} da_2 \
e^{n_1a_1} e^{n_2a_2} \psi({\bf a}),
\end{equation}
evaluated at integer points, ${\bf n} \in {\bf {\cal
Z}}^2$.

Unfortunately, ${\bf L}$ is not defined for
all symmetric ${\bf {\cal L}}^2({\cal R})$ functions\footnote{We
will refer to
the set of such symmetric ${\bf {\cal L}}^2$ functions as
${\bf {\cal L}}^2_+$.}.
Let ${\cal D}_{{\bf L}}$ be the domain in
${\bf {\cal L}}^2_+$ on which ${\cal {\bf L}}$ is defined.  Note that ${\cal
D}_{{\bf L}}$ is a linear space that is dense in ${\cal
{\cal L}}^2_+$ as it contains all $C^{\infty}$ functions of
compact support.

At this point, a theorem about inverting the
two-sided Laplace transform will prove useful to our
discussion.  From Ref. \cite{Widder}, we find that
if a holomorphic function
$\psi _* (s)$ defined on the complex plane satisfies
the following requirements:

\par

i) For each $s \in {\bf {\cal R}}$, there is some
$M_s \in {\bf {\cal R}}$ such  that $\Big| \int_{-\infty}^{\infty}{dt
 \psi _* (s + it)} \Big| <  M_s$,

\noindent and

ii) The limit, $\lim_{|t | \rightarrow \infty} \psi_*(s+ i t)$
converges uniformly to zero for $t$ in any

\ \ \ closed interval in the real line,

\par

\noindent then there
is some function, $\psi (a)$, such that $\psi _*$ is
the two-sided Laplace transform of $\psi$.  The two
functions are related by

\begin{equation}
\label{rel}
\psi (a) = \int_{-\infty}^{\infty} {dt \ \psi_*(it) \exp(-ita)} \quad \rm{and}
\quad
\psi_* (s) = \int_{-\infty}^{\infty} {da \ \psi(a) \exp(as)}
\end{equation}
Thus $\psi_*(it)$ is just the Fourier transform of $\psi (a)$
if the above conditions are satisfied.

We will continue to use this notation:  Bold arguments take values in a
two dimensional space while normal print
arguments take one dimensional values.  A function which is a candidate for
being the two-sided Laplace transform
of a function $\psi ({\bf a})$ of connections  will be denoted by a
*-subscript $\psi _*({\bf s})$.  In
addition, we will denote by $\tilde
{\psi}$ the Fourier transform of $\psi ({\bf a})$
so that $\tilde {\psi } (x) = \psi _* (ix)$.
A hat ($\hat {\psi}$) will
denote the potential image of a Laplace transform evaluated at integer
arguments; $\hat {\psi} (n) = \psi _* (n)$ for $n \in {\cal Z}$.

In trying to resolve our dilemma, we
might first ask if the image of the loop transform
($Im({\bf L})$) in the spacelike sector contains the
``standard" loop representation (the set of $l^2$
functions loops) at all.  For, if the loop
transform defines a distinct loop representation in the
spacelike sector by producing a representation containing no
function with support on a single loop,
then the proof in Appendix \ref{uip}
would not apply and
our dilemma would be resolved.  Are any such functions of finite
support in $Im({\bf L})$?  In fact, they all are, and we can
use the above theorem to show this.

Note that the
function
\begin{equation}
\label{charfuncts}
\psi _{*n}
(s) = e^{\alpha s^2} \Biggl( {{\sin{\pi (s-n)}}\over{\pi
(s-n)}} \Biggr) ^2 \end{equation}
takes on the values
\begin{equation} \hat
{\psi}_{n}(m) = \delta _{n,m} \end{equation}
at integer points.  Also note that this function
satisfies the conditions of
our transform inversion theorem for any $\alpha \in {\bf {\cal R}}^+$ and so
is the two-sided Laplace transform of some function $\psi(a)$.
Since the Fourier transform preserves the ${\cal L}^2$ norm, we can use
$\psi_{*n}(ix) = \tilde{\psi}(x)$ to evaluate $||\psi||_{{\cal L}^2}$ and
find that $\psi \in {\cal L}^2({\cal R})$.  It follows that if we define
$\Psi_{\bf n}({\bf a}) \equiv \psi_{n_1}(a_a) \psi_{n_2}(a_2) +
\psi_{n_1}(-a_1) \psi_{n_2}(-a_2)$, then $\Psi_{\bf n}$ is in ${\cal L}^2_+$
and
\begin{equation}
\hat{\Psi}_{\bf n}({\bf m}) = \delta_{{\bf n},{\bf m}} + \delta_{{\bf n},
{\bf -m}}
\end{equation}
and $\hat{\Psi}_{\bf n}$ is nonzero on exactly one distinct
loop.

Since the loop transform is a linear operator
and its image contains the functions  $\hat{\Psi}_{\bf n}$, all
functions with support on a finite set of loops are also in the
image of the loop transform.  The image of the loop transform therefore
contains at least a dense subset (in the $l^2$ topology) of the standard
loop representation.  A more subtle use of the transform
inversion theorem shows that
{\it every} $l^2$ function of loops is in the image of the
loop transform.  This fact is a straightforward consequence of the
main proof given in Appendix \ref{proof} and so will be established
at the end of this appendix.  Of course, the transform inversion
theorem can also be used
to show that $Im({\bf L})$ contains functions of loops that are
not in $l^2$ at all, but this only complicates our dilemma.

Instead of asking questions about the image of the loop
transform we might ask questions
about this transform's kernel.  A nontrivial kernel would
tell us that the loop transform is not a representation isomorphism in
the spacelike sector and would go a long way toward resolving our
dilemma.  We can again use the transform inversion theorem to construct
functions that are in the kernel of
the loop transform.  For example, consider the
function:

\begin{equation}
\label{kerex}
\psi _* (s) = \sin^2(\pi s) \ e^{\alpha s^2}
\end{equation}
for $\alpha > 0$.
Note that this function satisfies the conditions of the transform
inversion theorem and therefore is the the
two-sided Laplace transform of some $\psi (a)$.   Since the ${\cal L}^2$
norm is preserved by the Fourier transform, we see
from Eq. \ref{rel} that $\psi \in {\bf {\cal L}}^2({\cal R})$ and therefore
that the function ${\bf \Psi}({\bf a}) = \psi (a_1) \psi (a_2)$
is in the Hilbert space of the connection representation although its
loop transform is the zero function: $({\bf L} {\bf \Psi})(n_1,n_2) =
\psi _* (n_1) \psi _* (n_2) = 0 \text{ for } {\bf n} \in
{\bf {\cal Z}}^2$ \footnote{Again, thanks to Jorma Louko for pointing
out these constructions}.  We conclude that ${\bf L}$ has a
nontrivial kernel; it is not a representation isomorphism.

Let us now try to answer the question ``How bad is it?"
In other words, how big is the kernel of the loop transform and just
what is the relationship between the
connection and loop representations in the spacelike
sector?  A brief look at the discussion above indicates
that the kernel $K({\bf L})$ of ${\bf L}$ may be large
indeed since $\hat {\psi}$ samples only the integer
values of $\psi _*$.  However, not all functions
$\psi_*$ are the Laplace transform of some $\psi \in {\bf
{\cal L}}^2$ so this argument cannot be naively used to describe $K({\bf L})$.
In particular, if $\hat {\psi}(n) = {\bf
L} \psi(n)$ is defined for all $n \in {\bf
{\cal Z}}$, then the integral defining the Laplace
transform converges for all complex $n$ and does so uniformly on each
compact subset of the complex plane.
Because this integral converges uniformly in any compact region,
we can differentiate the
resulting function with respect to $n$ by interchanging
the orders of integration and differentiation.
Therefore, since $e^{an}$ is analytic in $n$ for all
complex $n$, $\psi_*(s)$ satisfies the Cauchy-Riemann
equations and is also analytic for all complex $s$.  This restriction that
$\psi_*$ be analytic is a rather strong one and could severly restrict
the functions in the kernel, making the kernel much smaller than we might have
anticipated.  In addition to this, there is still the restriction that
$\psi(a)$ be normalizable in ${\cal L}^2$.  A more
careful examination will be required to determine the status
of this kernel.

The most benign possibility is that $K({\bf L})$ is
finite dimensional.  In this case, we could reasonably
conclude that the loop representation contains ``almost
all" of the information in the connection representation
and attempt to supplement the loop representation with
an additional (discrete) degree of freedom.  However, this loop
representation would still contain all
functions with support on a finite number of loops (a
dense set in the standard representation) and, if the discrete label
does indeed remove the ambiguities associated with the loop
functions then the loop function that is always zero would again
represent the zero state, at least for some value of the discrete
index.  This loop representation must again be a
topologically irreducible representation of the ${\cal
T}^0$'s and ${\cal T}^1$'s since ${\bf {\cal L}}^2_+$ is and
it follows that the comments in Appendix \ref{uip} still show
that the loop representation and its inner
product are uniquely given by the standard loop
representation.  Thus, the loop representation would
still have to be identical to the
standard loop representation and our dilemma would remain unresolved.

However, this is not the case.  If the
function given in \ref{kerex} is multiplied by any
polynomial in $s$, it still satisfies the conditions of
the transform inversion theorem and still gives zero when evaluated
at any integer.  Since these functions are
linearly independent, $K({\bf L})$ must be infinite
dimensional.

The next most benign possibility is that $K({\bf
L})$ is closed as a subspace of ${\bf {\cal L}}^2_{+}$.
In this case we might still argue that the loop
representation captures an essential part of the
connection representation.  Then, the connection
representation ${\bf {\cal A}}$ would be related to the loop
representation, $L$, by ${\bf {\cal A}} \cong K({\bf
L}) \oplus {\bf {\cal A}}/K({\bf L})$ and ${\bf
{\cal A}}/K({\bf L}) \cong L$ so that ${\bf {\cal A}}
\cong K({\bf L}) \oplus L$.  Again, this loop
representation must be just the standard loop
representation.  However, it turns out that the kernel is not in fact
closed.  The proof of this
makes further use of the transform inversion theorem and is given
in Appendix \ref{proof}.  This appendix shows that the product of any two
functions $\psi_n$ defined in \ref{charfuncts} is in the
closure of $K({\bf L})$ and therefore that
$K({\bf L})$ is not closed.

A slight modification of this last proposal would be to
make the decomposition ${\bf {\cal A}} \cong \overline{K({\bf
L})} \oplus {\bf {\cal A}}/\overline{K({\bf
L})}$, where the overbar represents the closure as a
subspace of ${\bf {\cal L}}^2_+$.  However, $L \cong {\bf
{\cal A}}/\overline{K({\bf L})}$ may be nothing like
the standard loop representation since it might be that
for some $\hat{\phi} \in Im{\bf L}$, the elements of the connection
representation that map to $\hat{\psi}$ via the loop transform are all
contained in the closure of $K({\bf L})$. In this case, even
though $\hat{\phi} \in Im({\bf L})$, we would still find that
$\phi \equiv 0$ in
$A/\overline{K({\bf L})}$.

An important question is ``Which $\hat{\phi}$ are identified with the
zero state by this procedure?"
Or, ``How much of the connection representation does this
loop representation capture?"  These questions are related to the
question, ``How large is the kernel?" which is explored in Appendix A.  By
using the transform inversion theorem several times, this appendix shows
that the kernel is so large that for any symmetric or antisymmetric
function  $\Lambda$ in ${\cal L}^2({\cal R})$
of compact support, there is a sequence of ${\cal L}^2({\cal R})$
functions that
converges to $\Lambda$ in the
${\cal L}^2({\cal R})$ topology such that the
Laplace transforms of {\it each} function in this sequence
vanishes for all integer values.
Thus, the closure in this topology
of the set of ${\cal L}^2({\cal R})$ functions whose
Laplace transforms vanish at all integer values
contains all symmetric or antisymmetric ${\cal L}^2$ functions of
compact support on the
real line.  Since this closure is closed, it must
also contain the closure of the set of all symmetric and antisymmetric
${\cal L}^2$ functions of
compact support, which is just the space of all symmetric and
antisymmetric ${\cal L}^2$ functions.  A dense subset of  ${\cal L}^2_+$
consists of a sum of products of symmetric ${\cal L}^2({\cal R})$ functions
in each variable and
products of antisymmetric  ${\cal L}^2({\cal R})$ functions in each
variable.  It follows
that ${K({\bf L})}$ is dense in ${\cal L}^2_+$ and that ${\bf
{\cal L}}^2_+/\overline{K({\bf L})} \cong \{0 \}$!  Thus there is a
sense in which the loop
transform does not capture any part of
the connection representation.

This result is indeed strong enough to let us out of our dilemma.
If we look simply
at the image $Im({\bf L})$ as a loop representation then the zero function
does not necessarily represent the zero state.  If we look at the
quotient $Im({\cal L}^2)/\overline{K({\bf L})}$ then this
space does not contain any function of loops that is nonzerop on
exactly one loops.  In fact, it doesn't contain any non-zero
functions of loops at all.  Neither choice satisfies the conditions
required in Appendix \ref{uip} so that the proof given there has nothing
at all to say about the character of the resulting loop representation.

\section{Constructing a ``Loop" representation}
\label{looprep}

We have seen that there is a sense in which the loop transform does
not capture any part of the connection representation.
Is there another sense in which we can construct a ``loop
representation" that is in fact
isomorphic to the connection representation?
Despite the discussion of the last section, we
shall see that the answer to this question is, ``Yes.".
We begin by recalling that a loop representation is
a representation of the ${\cal T}^0$, ${\cal T}^1$ algebra on some
space which is the completion of some linear space of functions of
loops.  This means that we do not need to map every element of the
Hilbert space of the connection
representation to a function of loops, but only the elements of some
dense subset, ${\cal V}$.

The idea is to construct a loop representation by carefully choosing this
dense subset.  While $K({\bf L})$ is dense in ${\cal L}^2_+$,
suppose that it is also true that there exists some dense, linear
subspace, ${\bf
{\cal V}} \subset {\cal D}_{\bf L}$ and such that ${\bf {\cal V}} \cap
K({\bf L}) = \{0\}$\footnote{Recall that two dense subspaces {\it can} have
trivial intersection.}.  Let us also assume that ${\bf {\cal V}}$ is
stable under the ${\cal T}^0$'s and ${\cal T}^1$'s so that it
does, in fact, provide a representation of our basic
algebra.  Since ${\bf {\cal V}} \cap K({\bf L}) =
\{0\}$, it follows that ${\bf {\cal V}}$ and ${\bf L}({\bf {\cal V}})$
are isomorphic as representations of the ${\cal T}^0$'s and
${\cal T}^1$'s.  If we now define the
inner product on ${\bf L}({\cal V})$ by $\langle {\bf
L} \phi, {\bf L} \psi \rangle \equiv \langle \phi , \psi
\rangle$, then ${\cal V}$ and ${\bf L}({\cal V})$ are
isomorphic as vector spaces with an inner product.  Both of these spaces
may be completed in the topologies defined by their inner products
and the results must be isomorphic as Hilbert spaces.

This Hilbert space isomorphism can now be used to define the action of the
${\cal T}^0$'s and ${\cal T}^1$'s in the completion of ${\bf L}({\cal V})$ in
such a way that this completion $\overline{{\bf L} ({\cal V})}$ and
the completion of ${\cal V}$ are isomorphic as
representations of the ${\cal T}^0$, ${\cal T}^1$ algebra.  But, as the
completion of ${\cal V}$ is ${\cal L}^2_+$, our ``loop
representation"\footnote{The closure of ${\bf L}({\cal V})$ carries
a representation of the loop algebra and therefore defines a loop
representation.} is
isomorphic to ${\cal L}^2_+$ as a Hilbert space representation of
the ${\cal T}^0$, ${\cal T}^1$ algebra.  If such a ${\cal V}$ exists, we see
that
it defines a loop representation which is isomorphic to the connection
representation, although we had to take special care to construct it.

The construction of such subspaces and the proof that they satisfy our
requirements requires additional care, so we carry it out in Appendix
\ref{Vs}.  In fact,
in Appendix \ref{Vs}  we show that there are two such subspaces.  Note that the
preceding paragraph shows that the representation constructed from
either subspace is isomorphic to the connection representation and
therefore that these two representations are isomorphic to each other.
Therefore, the prescription given above defines a unique loop
representation up to representation isomorphism.

Despite the uniqueness of this loop representation in the sense just
mentioned, there are still ambiguities in the formulation of these
representations in terms of loops.  We should note that
the above procedure constructs loop representations only in a
technical sense.  That is, they are completions of a
linear space of functions of loops, but may therefore contain many
elements that cannot, in fact be described as a function of loops.

We should also be aware that even though these loop representations are
isomorphic, this does not mean that the isomorphism acts in a way
that preserves the
parts of these representations which are formed by functions of loops.
Nor does it mean
that functions of loops have a well defined correspondence with functions
of connections without reference to the specific subspace ${\cal V}$ used to
construct that loop representation.  We shall see that a given function of
loops may correspond to a different function of connections depending
on which subspace was chosen.  These functions of connections may
even have different norms so that the inner product on functions of
loops also depends on the subspace ${\cal V}$ that was chosen.
In fact, which functions of connections
correspond to functions of loops at all will also depend on
the choice of a dense subspace ${\cal V}$.

To see why these ambiguities arise, let's try
the above construction with two subspaces, ${\bf {\cal V}}_1$, ${\bf
{\cal V}}_2$, both
satisfying the conditions given above.  First, we notice
that for any $\hat{\phi} \in
Im({\bf L})$, there is a dense subspace of the carrier space of the
connection representation that maps onto $\hat{\phi}$.  Since only one of
these can be represented by $\hat{\phi}$, only a rather small part of this
carrier space will correspond to functions of loops.
Just which $\phi$ will be
so represented depends on the choice of ${\bf {\cal
V}}$; if there is some $\phi_1 \in
{\bf {\cal V}}_1$ and some
$\phi_2 \in {\bf {\cal V}}_2$
such that ${\bf L} \phi_1 = \hat{\phi}=
{\bf L} \phi_2$, $\hat{\phi}$ would represent $\phi_1$ in the representation
constructed from ${\cal V}_1$ and would represent $\phi_2$ in the
representation constructed from ${\cal V}_2$.  Since the
function $\hat{\phi}$ already represents $\phi_1$ in the first representation
there is no way for $\phi_2$ to be represented there by a
function of loops consistent with the loop transform.  If
$||\phi_1|| \neq ||\phi_2||$, then
$||\hat{\phi}||$ also depends on which ${\bf {\cal
V}}$ we used as a starting point and therefore so does the
form of the inner product in terms of functions of loops.
We show in
Appendix \ref{Vs} that
${\bf {\cal V}}_1$ and ${\bf {\cal V}}_2$ as discussed
above do indeed exist.  Therefore, we cannot hope that the
requirements imposed on ${\cal V}$ conspire to make the choice of ${\bf {\cal
V}}$ unique or to remove the above ambiguities.

However, all known subspaces ${\bf {\cal
V}}$ that satisfy the
above requirements are highly algebraically reducible
representations of the ${\cal T}^0$'s and ${\cal T}^1$'s.
They have the property that every linear
subrepresentation of the ${\cal T}^0$'s and ${\cal T}^1$'s is
also algebraically reducible. It would be fairly natural
to require the subspace ${\bf {\cal V}}$ that we will
use to construct a loop representation to be an
algebraically irreducible representation of the ${\cal
T}^0$'s and ${\cal T}^1$'s and it might be hoped that such a
requirement would select a unique subspace
or at least remove the above mentioned ambiguities
in the loop representation.  At present, it is not known
if any such subspaces of ${\bf {\cal L}}^2_+$ exist.  If
they do, we could then explore the question of uniqueness.

\section{Generalized Loop Coordinates and the inner
Product}
\label{ip}

While we have been unable to remove these ambiguities in our loop
representation, there {\it is} a way to construct a related but different
representation that is not subject to such ambiguites.  This
representation will not be formulated in terms of functions of loops, but in
terms of functions of ``generalized" loops.  Before we construct this
representation, we will briefly review the definitions of generalized
loops and the corresponding generalized holonomies from
\cite{SEL,LSC} and a few of the relevant facts derived in those papers.
We will then proceed to construct a generalized loop representation
using suitable functions of generalized loops and generalized loop
analogues of the ${\cal T}^0$'s and ${\cal T}^1$ operators.  This
construction will again rely heavily on a transform which is
a generalized version of the loop transform.

We now abandon the practice of sections \ref{QT}-\ref{looprep} of
using ``loops" to mean equitopy
classes of loops.  In this section ``loop" will refer to an equivalence
class under reparameterization of smooth
maps from [0,1] to some two manifold $\Sigma$ that map both 0 and 1 to
some fixed base point, *, in the manifold.  We also identify two such
maps $\alpha$ and $\beta$ if the composition of based loops $\alpha \circ
\beta^{-1}$, where $\beta^{-1}$ is the loop $\beta$ followed in the reversed
direction, is contractible within itself to the trivial loop.
These equivalence classes of based loops will be the starting point for
our description of generalized loops.

The idea behind generalized loops, also called extended loops and
loop coordinates, is to generalize loops in such a way as to form a space
that has the local structure of a Lie group \cite{SEL,ELG}.  Note that
while the space of loops as defined above has a natural group structure
in which multiplication of two loops is given by the composition of
pointed loops at their base point, this
is not a Lie group structure since there are no
one-parameter subgroups.  For example, if $\alpha$ is injective as a map into
its image then there is no loop that gives $\alpha$ when squared.  Roughly
speaking then, the ``extended group of loops" contains elements that ``fill in"
between the zero loop and once around a given loop.  It also contains elements
that fill in between once around a given loop and twice around that loop.
Suppose that we label this
one-parameter subgroup of extended loops $l_{\lambda}$ by
a real number $\lambda$.  For integer $\lambda$, $l_{\lambda}$ is the
set of loop coordinates of the loop that wraps $\lambda$ times around
our given loop and for all real $\lambda_1$ and $\lambda_2$ these labels
satisfy $l_{\lambda_1} \times l_{\lambda_2} = l_{\lambda_1 + \lambda_2}$,
where $\times$ is the group multiplication.
We might therefore say that these extended loops correspond to a ``loop"
that wraps $\lambda$ times around our given loop even for real $\lambda$.
We might also guess that the extension in this
sense of our set of equitopy classes of loops ${\cal Z}^2/{\cal Z}_2$ on
the torus is just ${\cal R}^2$ or perhaps ${\cal R}^2/{\cal Z}_2$.  This
is in fact the correct result, though we will need to introduce some of the
formalism of \cite{SEL,LSC} to show this explicitly.

For convenience, we will not introduce the full formalism of loop
coordinates since it is unnecessary for our simple model.  In particular,
we will only introduce enouugh formalism to describe generalized loops
for flat abelian connections.  This will suffice for our purposed since,
in the $\epsilon_1 = +1 = \epsilon_2$ spacelike sector,
every flat connection on the torus is gauge equivalent under small gauge
transformations to some
homogeneous connection.  This is easy to check since the
$SU(1,1)$ elements $\exp{[A_b^I \tau_I]}$ for two parallel 2+1 dimensional
internal Minkowsi vectors ${\bf A_b}$ are the holonomies corresponding to
a single loop in the $x_b$ direction given by the homogeneous
connection whose components are just this
same set of numbers $A_b^I$ at each point on the torus.  Here we are
using a standard set of coordinates $(x^1, x^2)$ on the torus with unit
period.  Such a connection behaves much like an abelian
connection since
the connection evaluated at any point on the torus and contracted with some
tangent vector to the torus at that point commutes with the
connection evaluated at any other point and contracted with any tangent
vector at this second point.

The holonomies of an abelian connection are simply exponentials of integrals
over each loop:
\begin{eqnarray}
U[\alpha](A) &=& \exp{\bigl( \int_{\alpha} dx^a \ A^I_a \tau_I \bigr)} \cr
&=& \exp {\Bigl( \int d^2x \ A^I_a \tau_I \bigl( \int_0^1ds \dot{\alpha}^a(s)
\delta^2(\alpha(s)-x) \bigr) \Bigr)}
\end{eqnarray}
The distribution: $ \int_0^1ds \dot{\alpha}^a(s)
\delta^2(\alpha(s)-x) \equiv T^a_{\alpha}(x)$
forms the set of so-called ``loop coordinates"
for the loop $\alpha$.  Note that this distribution satisfies:
\begin{equation}
\label{ff}
\partial_a T^a_{\alpha}(x) = 0
\end{equation}
for any loop $\alpha$.  Note that $T_{n \alpha}^a (x) = n T_{\alpha}^a (x)$.
Thus, one way to ``fill in" the gaps between the loops is to consider
any ``form factor" that satisfies Eq. \ref{ff} to define a generalized loop.

Corresponding to any such generalized loop, we can define a ``generalized
holonomy" by
\begin{equation}
U_T(A) = \exp{\bigl( \int d^x \ T^a(x) A_a^I(x) \tau_I \bigr)}
\end{equation}
Note that the condition (\ref{ff}) on the form factors guarrantees
gauge invariance of the holonomies (since the connections are abelian).
Juust as for loops, we will place the generalized loops into equivalence
classes so that two generalized loops are equivalent if they give the
traces of their generalized holonomies are the same for every connection
$A$ in whatever class of connections we are considering.

Let us now
compute the generalized holonomies that correspond to the form factors
that satisfy Eq. \ref{ff}.  Note that this equation implies that
 $T^a(x)$ must be of the form: $T^a(x) = \epsilon^{ab} \partial_b
\Lambda(x) + s^a$ for some function $\Lambda$ and some constant vector $s^a$.
We can use our
homogeneous connections to calculate these generalized holonomies so that:
\begin{eqnarray}
U_T(A) &=& \exp{[\int_{{\cal T}^2} d^2x \ A_b^I(x) T^b(x) \tau_I]} \cr
&=& \exp{[\int_{{\cal T}^2} d^2x (\epsilon^{bc} \partial_c \Lambda A_b^I)
\tau_I + A_b^I \tau_I s^b]} = \exp{(A_b^I s^b \tau_I)}
\end{eqnarray}
and distinct equivalence classes of generalized loops depend
only on the vector ${\bf s} = (s^1,s^2)$.

Because these generalized holonomies are boosts, each
choice of ${\bf s} \in {\cal R}^2$ leads to a distinct set of generalized
holonomies.  However, ${\bf s}$ and $-{\bf s}$ provide the same set
of traces.  Thus, we make one last identification of ${\bf s}$ with
$-{\bf s}$ and find that our equivalence classes are labeled exactly
by ${\bf s} \in {\cal R}^2/{\cal Z}_2$, just as we expected.

To complete our generalized loop representation, we would like
to form ${\cal T}^0$'s and ${\cal T}^1$'s
for each equivalence class ${\bf s}$ of generalized loops.  This can be
done by simply replacing
every holonomy in the definitions of these functions given in Eq.
\ref{connectionts} with a generalized holonomy.  Note that
the ${\cal T}^1$'s
are again well-defined as functions of our equivalence classes of
generalized loops.

These functions of generalized loops also satisfy the Poisson Bracket
relations given in Eq. \ref{comms} and the algebraic relations given in
Eq. \ref{allalgrels}
although the generalized loop
indices are now allowed to be real.  These functions can be expressed
explicitly in the connection representation as:

\begin{mathletters}
\begin{equation}
\label{genloopt0}
{\cal
T}^0[{\bf s}]({\bf a}) = \cosh ({\bf s} \cdot {\bf a})
\end{equation}
\begin{equation}
{\cal T}^1[{\bf s}]({\bf a}) = \sinh ({\bf s} \cdot {\bf
a}) \ {\bf s} \times {\bf p}
\end{equation}
\end{mathletters}
and in the generalized loop representation as:

\begin{mathletters}
\begin{equation}
({\cal T}^0({\bf t}) \psi)({\bf s}) = (\psi({\bf s} +
{\bf t}) + \psi({\bf s} - {\bf t}))/2
\end{equation}
\begin{equation}
({\cal T}^1({\bf t}) \psi)({\bf s}) = i {\bf t}
\times {\bf s} \ (\psi({\bf s} + {\bf t}) - \psi({\bf s} -
{\bf t}))
\end{equation}
\end{mathletters}

We can also define a generalized loop transform by
integrating any function in the connection representation
against traces of the generalized holonomies.  Thus, we evaluate the
loop transform integral not just for traces of holonomies around loops,
but also for traces of generalized holonomies of generalized loops.
{}From Eq. \ref{genloopt0}, we see that the loop
transform is again the two-sided Laplace transform, but
it is now evaluated at every ${\bf s} \in {\bf
{\cal R}}^2$.  Note that if the Laplace transform exists for some $s_1,
s_2 \in {\bf {\cal R}}$, then it exists for all $s_1 < s <
s_2$ and that therefore the existence of the generalized loop
transform for all real $s$ is equivalent to the existence
of the loop transform for all integer $n$ and the domain
of this generalized loop transform in the connection representation
is the same as the
domain ${\cal D}_{\bf L}$ of the original loop transform.

The generalized loop representation
avoids the ambiguities that we found for the original loop representation
because it has trivial kernel.  To see this,
recall that if the Loop transform
integral converges for all real ${\bf s}$, then $\psi_*({\bf s}) = \int d^2a\
e^{{\bf s \cdot a}} \psi ({\bf a})$ for all complex ${\bf s}$.  But,
if $\psi \in
K({\bf L})$, $\psi_*({\bf s}) = 0$ for all complex ${\bf s}$ and
$\tilde{\psi}({\bf x}) = 0$ as well.  But $\tilde{\psi}({\bf x})$ is
just the Fourier transform of $\psi({\bf a})$ and the Fourier transform has
trivial kernel in ${\cal L}^2_+$.  It follows that $\psi({\bf a}) = 0$
and that the kernel of the generalized loop transform is trivial.

This delightful fact means that we may construct our generalized
loop representation in much the same way that we built our loop representations
in section \ref{looprep} but that we may now use all of ${\cal D}_{\bf L}$
as the dense subspace whose image we complete to define the Hilbert space.
As before, we define the inner product in the image of
${\bf {\cal D}}_{{\bf L}}$ to be given by the inner product
in ${\bf {\cal D}}_{{\bf L}}$ and complete the
resulting space with respect to this inner product.  This produces
an unambiguous representation in terms of
(generalized) functions of generalized loops that is isomorphic to
the connection representation.  We must
consider suitably generalized functions since the
generalized loop transform is still not defined for all
$\psi \in {\bf {\cal L}}^2_+$ and the completion of $Im({\cal D}_{\bf L})$
will contain elements that are not functions of generalized loops.
The representation is
unambiguous in the sense that the part of the
representation that can be described by functions of
loops is exactly the same as what would have been obtained if any
dense subset ${\cal V}$ of ${\bf {\cal D}}_{{\bf L}}$ were used in the
construction instead
of ${\bf {\cal D}}_{{\bf L}}$ itself.

Similarly, we conclude that a given function of
generalized loops (in the image of ${\bf {\cal D}}_{\bf
 L}$) has a unique norm and a unique inner product with any
other such function of generalized loops.  We can even find an explicit
form for this inner product using our transform inversion theorem.
Remembering that our functions in the image of ${\bf {\cal
D}}_{\bf L}$ are holomorphic on the complex two-plane ${\cal C}^2$,
they have a unique and well defined
analytic continuation to the imaginary axes.  The inner product is
given in terms of this analytic continuation by:

\begin{equation}
\label{sttrouble}
\langle \phi , \psi \rangle = \int{d^2x \
({\phi}_*(i{\bf x}))^{\star}{\psi}_*(i{\bf x})}
\end{equation}
where ($\star$) represents complex conjugation and (*)
as usual denotes a function of generalized
loops.

The completion of the image of ${\bf {\cal D}}_{\bf
L}$ can be described easily in terms of functions of
the imaginary coordinate ${\bf x} = i{\bf s}$ as the ${\bf {\cal
L}}^2_+({\cal R}^2)$ space.  However, an attempt to describe this space
completely in
terms of functions of generalized loops (the real
axis) involves a prescription for analytic continuation
of an arbitrary ${\bf {\cal L}}^2({\cal R}^2)$ function defined along the
imaginary axes to the real axes.  Some sort of additional structure would
have to be introduced for this purpose.

On the other hand, suppose for a moment that we wish to change our position
and think about functions of ``imaginary generalized
loops" (i.e., to work with functions defined along the
imaginary axis).  While this provides a representation that
is clearly equivalent to the original connection
representation and in which it is easy to compute inner
products, the action of the ${\cal T}^0$'s and
${\cal T}^1$'s in this representation is far from simple.
The ${\cal T}^0$'s and ${\cal T}^1$'s are then  represented by
non-local differential operators, although the original
${\bf a}$ and ${\bf p}$ are represented simply.

We therefore see that the use of generalized loops eliminates
all of the ambiguities that we found in section \ref{looprep}.  However,
it is still true that not all states in the generalized loop
representation can be regarded as functions even of generalized loops.
Interestingly, they can be regarded as functions of ``imaginary
generalized  loops" and the inner product can be expressed conveniently
using this description, though this is at the cost of loosing
the simple form of the ${\cal T}^0$'s and ${\cal T}^1$ operators.  In
the end, this representation in terms of imaginary generalized
loops is just the momentum representation obtained by Fourier transform
of the connection representation.

\section{Discussion}
\label{Diss}

In thinking about this loop transform, one
analogy has proved itself to be extremely useful.
This is an analogy with the so-called (integer) ``Mellin transform,"

\begin{equation}
\psi_*(n) \equiv \int_{-\infty}^{\infty} x^n \psi (x) dx
\end{equation}
for $\psi \in {\cal L}^2({\cal R})$.  Here, the multiplication
operator $x^n$ plays the role of the ${\cal T}^0({\bf k})$'s and
the differential operators $-ix^n \case {\partial} {\partial x}$
play the part of the ${\cal T}^1({\bf k})$'s.  These
operators have a commutator algebra reminiscent of the
loop algebra and the set of polynomials in $x$ provides a
cyclic representation for these operators.

If we consider the action of the Mellin transform in the momentum
representation:

\begin{equation}
\psi(n) = \int x^n \tilde{\psi}(p) e^{-ixp}dxdp = 2 \pi \bigl( -i \case
{\partial} {\partial p}  \bigr) ^n \tilde{\psi}(p) \bigg|_{p=0}
\end{equation}
we see that the resulting function of integers determines
$\psi$ uniquely only if we require that $\tilde{\psi}(p)$ be
analytic.  All functions of momenta that are zero in some neighborhood of
the origin are in the kernel of this transform and this set is dense in
${\cal L}^2({\cal R})$.  Again, several
dense linear subspaces of ${\cal L}^2$ can
be found that provide a representation of the above algebra but don't
intersect the kernel of the transform and again the images of these subspaces
overlap.  This analogy may continue to be useful in future study
of the loop transform.

Despite the dense kernel of the loop transform,
we did find that a loop representation
can be constructed in the spacelike sector that is isomorphic to
the connection representation.  The construction of this loop
representation is based on the choice of some dense subspace ${\cal V}$
of the
connection representation satisfying certain conditions.  Because
the construction process guarantees that the loop representation
is isomorphic to the connection representation, the loop
representation constructed using any choice of suitable dense subspace is
isomorphic to that constructed using any other such choice.  However,
a number of ambiguities remain concerning just how this representation
is described in terms of loops and it is difficult to see how the
loops themselves could be considered to be the fundamental
objects in the theory.  It is possible that
requiring the dense subspace space ${\cal V}$ to be an algebraically
irreducible representation of the ${\cal T}^0$'s and ${\cal T}^1$'s will
remove these ambiguities.  However, at present it is not even known
if such a subspace exists.

While we have confined ourselves to 2+1 gravity on a torus,
we might try to use our results to draw more general conclusions.
There are, however, several potential obstacles to doing so.  One potential
obstacle is that 2+1 dimensional gravity on a torus is an
abelian theory, in contrast to 3+1 gravity or even 2+1
gravity on higher genus spatial slices.  Another potential obstacle is
that we have not considered the possibility of imposing
identifications in the phase space under
large diffeomorphisms.  To impose such identifications, we would
have to face
two problems: first, that the large diffeomorphisms do not act
properly discontinuously on our reduced phase space; second, that
the ${\cal T}^0$'s and ${\cal T}^1$'s are not invariant under large
diffeomorphisms.  This second problem means
that special consideration would be
required to define the loop transform if we quotient the reduced
phase space by the large diffeomorphisms.

Despite these potential obstacles,
let us briefly speculate about 2+1 dimensional
gravity on $\Sigma \times {\cal R}$, where $\Sigma$
is a compact surface with genus greater than one.
Note that the reduced configuration space (without dividing by
the large diffeomorphisms) has infinite volume\footnote{This follows from
a result in \cite{inf}.  Thanks to Steve Carlip for help finding this
reference.} and that the
holonomies are boosts in the component of the phase space that
corresponds to classical spacetimes with no closed timelike curves\cite{Mess}.
Because of this,
the ${\cal T}^0({\bf k})$'s are unbounded and we would expect results
similar to the spacelike sector of the torus.  It should be emphasized,
however, that at
this point this is only speculation by analogy.

We might also try to draw conclusions
about loop representations in 3+1 dimensional gravity.
However, the 3+1 theory is extremely different,
not least because the ${\cal T}^0$'s and ${\cal T}^1$'s
are no longer classically real and so cannot be taken to be Hermitian
in the quantum theory.  This means that the inner product must be determined
in a different way in the 3+1 theory.  Since it is the inner product
that determines the continuity of the loop transform, this makes it
unclear what implications our work holds for the 3+1 theory.

However, when trying to use our results to discuss the 3+1 theory,
the fact that we did not take the
quotient of the reduced phase space by the large diffeomorphisms is
actually not an obstacle, but an advantage.  2+1
gravity on a torus is a theory with a finite number of degrees
of freedom (2), describing only the global modes of a 2+1
spacetime and the rearrangement of our set of loops under
large diffeomorphisms is a direct consequence of their global nature and the
fact that they represent homotopy classes.
On the other hand, ``loops" in the 3+1 theory represent not
homotopy classes of loops, but much smaller classes of loops
that only contain loops that differ from each other by retracings \cite{Ash}.
This is
related to the fact that 3+1 gravity is a field theory with both local
and global degrees of freedom.  Given two independent holonomies (loops)
in the 3+1 theory, there is no diffeomorphism (large or small) that
maps these loops to products of themselves as the large diffeomorphisms
map the loops in
the 2+1 theory.  We may therefore hope that we have chosen a description of
2+1 gravity that can teach us about the 3+1 theory.

One general lesson that we can claim to have
learned here is to be wary of formal arguments
using the loop transform.  In general, extra structure
may be required in order to give the loop representation the
desired properties.  The resulting loop representation
may be nothing like what might have been expected by
thinking only about loops.  Indeed, if the goal is
equivalence between the connection and loop representations -- or, more
generally, to ensure that the loop representation provides the correct
classical limit  --
then it is difficult to see what conceptual role the loops
themselves take on.

On the other hand, generalized loops brought
us much more success.  Though the inner
product on the space of functions of generalized loops
is still quite complicated, at least it is well defined.
Because generalized loops form
a vector space and lack the discrete structure of loops, they are
easier to link to a connection representation.  However, this same feature
can make them less appealing to those who would like to see
a discrete structure for spacetime.  Constructions analogous to
the weave states and to various operators that have been defined in
loop representations \cite{Smo,trans,weave} may have to be modified
and the conclusions rethought if they are reformulated in terms of
generalized loops.
Yet, it seems likely
that generalized loops will continue to play a significant role in
the study of loop representations.

On the whole, we have seen that the
construction of a loop representation in the spacelike
sector of 2+1 quantum gravity is a subtle
issue.  This is a consequence of the fact that the loop
transform in the spacelike sector is not everywhere
defined on the connection Hilbert space fails to be
continuous with respect to the natural topologies, even in its
domain of definition.  Any attempt to define the loop
representation by completion of all or part of the
image of the domain of definition of the loop
transform must somehow introduce some extra structure to
deal the the issues raised by
lack of continuity.

\acknowledgements
The author would like to express his thanks to Abhay Ashtekar and
Jorma Louko for numerous questions and comments.  Special thanks go to
Abhay Ashtekar for editorial comments, to Alan Rendall
for punctuating the difference between algebraic and
topological reducibility, and to Jorma Louko for asking the question
``Is the kernel dense?"  This work was supported in part by the NSF grant
PHY XXX to Syracuse University and by research funds provided by
Syracuse University.

\appendix
\section{The Kernel is Dense}
\label{proof}

In order to show that the
kernel $K({\bf L})$ is dense in ${\bf {\cal L}}^2_+$, we
will show that for each continuous $\Lambda \in {\bf {\cal L}}^2_+$ of
compact support that is invariant under ${\bf a} \rightarrow -
{\bf a}$\footnote{We will refer to this set of functions
as $C^0_{0+}({\bf {\cal R}}^2)$}
there is a sequence, $\{ \phi^{\Lambda,(k)} \} \subset
K({\bf L})$ such that $\phi^{\Lambda,(k)} \rightarrow
\Lambda$ in the ${\bf {\cal L}}^2$ norm as $k \rightarrow
\infty$.  From this, it will follow that the closure $\overline{K({\bf
L})}$ of the kernel
contains ${\cal C}_{0+}^0({\bf {\cal R}}^2)$ and that therefore we
have $\overline{K({\bf L})} \supset \overline{{\cal
C}_{0+}^0({\bf {\cal R}}^2)} \supset {\bf {\cal L}}^2_+$.  We conclude that
$K({\bf L})$ is dense in ${\bf {\cal L}}^2_+({\bf
{\cal R}}^2)$.

The sequence, $\{ \phi^{\Lambda,(k)} \}$,
will be constructed by first finding a sequence of
functions, $\{ \psi^{\Lambda,(k)} \}$, that have the same
loop components as $\Lambda$:

\begin{equation}
\hat{\psi}^{\Lambda,(k)}({\bf n}) = \hat{\Lambda}({\bf n})
\end{equation}
This sequence will also have the property
that $\{ \psi^{\Lambda,(k)} \} \rightarrow 0$ as $k
\rightarrow \infty$.  This sequence will in turn be
constructed from a sequence $\{\psi^{(k)}_{{\bf
n}}\}$ of sets of
``characteristic functions of loops"
which have the properties that $\hat{\psi}^{(k)}_{{\bf
n}}({\bf m}) = \delta_{{\bf n},{\bf m}}$\footnote{The symbols ${\bf n}$ and
${\bf m}$ are taken to label equivalence classes of loops, i.e. ${\bf n},
{\bf m} \in {\cal R}^2/{\cal Z}^2$.} and that
$\psi^k_{{\bf n}}\ \rightarrow 0$ in ${\bf {\cal L}}^2_+$
as $k \rightarrow \infty$.  We then define $\psi^{\Lambda , (k)} \equiv
\sum^{\infty}_{m = 0} \hat{\Lambda}({\bf m}) \psi^{(k)}_{{\bf
m}}$ so that $\hat{\psi}^{\Lambda , (k)}({\bf n}) =
\hat{\Lambda}({\bf n})$.  The function $\phi^{\Lambda , (k)}$ is
defined by $\phi^{\Lambda, (k)} \equiv \Lambda -
\psi^{\Lambda , (k)}$.  The proof consists of showing the
existence of the functions $\psi^{(k)}_{{\bf n}}$ and that
the above limits behave appropriately.  Actually, we do all
of this in the (completely) symmetric and antisymmetric
parts (${\bf {\cal L}}^2_{++}$ and ${\bf {\cal
L}}^2_{--}$) separately so that we can work with functions
with one-dimensional arguments.  The proof for our functions of
two-dimensional arguments follows by taking tensor products of
function spaces.

We begin by recalling that the loop transform is
just the two sided Laplace transform in each variable.
Therefore, if we can write our space of functions of connections
in an appropriate way in terms of functions of one-dimensional
arguments, we can work with the Laplace transform separately in
each argument.  To this end, we split the connection
representation into completely symmetric and completely
antisymmetric pieces, ${\bf {\cal L}}^2_+ = {\bf {\cal
L}}^2_{++} \oplus {\bf {\cal L}}^2_{--}$.  The space ${\cal L}^2_{++}$
consists of functions that are symmetric in $a_1$ and $a_2$ separately
and the space ${\cal L}^2_{--}$ contains those functions that are
antisymmetric in both coordinates.  A
dense subset of ${\cal L}^2_{++}$ (${\cal L}^2_{--}$) consists of
sums of products of symmetric (antisymmetric) functions
with one-dimensional arguments:

\begin{equation}
{\cal L}^2_{++}({\cal R}^2) = {\cal L}^2_+({\cal R}) \otimes {\cal L}^2_+
({\cal R}) \text{\ and } {\cal L}^2_{++}({\cal R}^2)
= {\cal L}^2_+({\cal R}) \otimes {\cal L}^2_+
({\cal R})
\end{equation}
where the $\pm$ subscripts on ${\cal L}^2({\cal R})$ again denote
the symmetric and antisymmetric parts.
Thus, if we can show that
the kernel of the two-sided Laplace transform evaluated at
integer points is dense in ${\bf {\cal L}}^2_{+}({\bf
{\cal R}})$ and  ${\bf {\cal L}}^2_{-}({\bf
{\cal R}})$, it will follow that the kernel of the loop
transform is dense in
${\bf {\cal L}}^2_+({\bf {\cal R}}^2)$.

For convenience, we introduce the symbol ${\bf L}^1$ to denote this
two-sided Laplace transform of a functions of a one-dimensional
argument; that is,
\begin{equation}
\hat{\psi}(n)  = [{\bf L}^1 \psi](n) = \int_{-\infty}^{\infty} da\ e^{na}
\psi (a)
\end{equation}
for all $n \in {\cal Z}$.  We also define $K_{\pm}({\bf L}^1)$ and
${\cal D}_{\pm {\bf L}^1}$ as subsets of ${\cal L}^2_{\pm}({\cal R})$
by analogy with $K({\bf L})$ and ${\cal D}_{\bf L}$.

A fundamental part of our proof will be
the ``characteristic functions of loops" defined in
Eq.  \ref{charfuncts}. We now define the symmetrized and
antisymmetrized combinations:

\begin{equation}
\psi_{*n}^{\pm (k)} (s) = \Big( \psi_{*n}(s) \pm
\psi_{*-n}(s) \Big)^k \ \ \rm{ for } \ n \neq 0
\end{equation}
and

\begin{equation}
\psi_{*0}^{+ (k)} (s) = \Big( \psi_{*0}(s)\Big)^k
\end{equation}
where we always take $k$ odd so that $\psi_{*n}^{+(k)}
(s)$ is a symmetric function of $s$ and $\psi_{*n}^{-(k)}(s)$
is an antisymmetric function of $s$.  Note that,
$\psi_{*n}^{\pm (k)} (m) = \delta_{n,m}$ for $n,m \in {\bf
{\cal Z}}^+ \cup \{0\}$.

Since each of the above functions is analytic on all
${\bf {\cal C}}$, is integrable in absolute value along
every contour parallel to the imaginary axis, and
uniformly converges to zero in the imaginary direction
for $\Re (s) \in [a,b]$ for any closed interval, $[a,b]
\subset {\bf {\cal R}}$, these functions satisfy the
conditions of the transform inversion theorem.  There is
therefore
some $\psi^{\pm (k)}_{n}(a)$ such that $\psi^{\pm (k)}_{*n}(s)$
is the Laplace transform of $\psi^{\pm (k)}_{n}(a)$.

In fact, $\psi^{(k)}_{n} \in {\bf {\cal
L}}^2_{\pm}({\cal R})$ since

\begin{equation}
||\psi^{\pm (k)}_{n}||^2_{{\bf {\cal L}}^2} =\int_{-\infty}^{\infty} {da
|\psi^{\pm (k)}_{n}(a)|^2} = \int_{-\infty}^{\infty} {dx |\tilde {\psi}^{\pm
(k)}_{n}(x)|^2} = \int_{-\infty}^{\infty} {dx |\psi^{\pm (k)}_{*n}(-ix)|^2}
\end{equation}
We can estimate these norms by using the fact that for
$n \geq 1$ and all real $x$,

\begin{equation}
|\tilde{\psi}^{\pm (k)}_{n}(x)| \leq e^{- k \alpha n^2}
[Q(x)]^k
\end{equation}
where for convenience we have set

\begin{equation}
Q(x) = 2 e^{- \alpha x^2} e^{2 \pi |x|}/ \pi^2 \leq
{{2e^{\pi^2/ \alpha}} \over {\pi^2}}
\end{equation}

Let's choose $\alpha > \pi^2$ so that $|\tilde{\psi}^{\pm
(k)}_{n}| \leq e^{- \alpha n^2} < 1$.  Note that

\begin{equation}
||Q^k||_{{\bf {\cal L}}^2} < {{8^k} \over {\pi ^{4k}}}
{{e^{2 \pi ^2 /k\alpha} \sqrt{\pi}} \over {\sqrt{2 \alpha
k}}} < {{1} \over {\sqrt k}}
\end{equation}
and $\psi^{\pm (k)}_{n} \rightarrow 0$ as $k \rightarrow
\infty$ as promised.  For $n=0$, we make use of the facts
that

\begin{equation}
\eqnum{fact i}
\label{facti}
{{\sinh {y}} \over {y}} \leq {{e^{|y|}} \over {1 + |y|}}
\end{equation}
and

\begin{equation}
\eqnum{fact ii}
\label{factii}
{{e^{\pi |y|}} \over {1 + \pi |y|}} \leq e^{\pi^2 |y|^2}
\end{equation}
for all real $y$.  These show that, for all real $x$,

\begin{equation}
|\tilde{\psi}^{+(k)}_{0}(x)| \leq e^{- k (\alpha - \pi ^2)
x^2}
\end{equation}
and therefore that $||\psi^{+(k)}_{0}||^2_{{\bf {\cal
L}}^2} < \sqrt{{{\pi} \over {2k(\alpha - \pi^2)}}}$ and
$\psi^{+(k)}_{0} \rightarrow 0$ as $k \rightarrow
\infty$.  For further convenience, let's choose
$\alpha > \pi^2 + 1$ so that we may write simply

\begin{equation}
||\psi^{\pm (k)}_{n}||^2_{{\bf {\cal L}}^2} < {{
e^{-2 \alpha n^2 k}} \over {\sqrt{k}}}
\end{equation}
for all $n \in {\bf {\cal Z}}^+ \cup \{0\}$.

Note that this is already enough to show that
$\psi^{\pm (1)}_{n} \in \overline{K_{\pm}({\bf L}^1)}$,
since $\psi^{\pm (1)}_{n} - \psi^{\pm (k)}_{n} \rightarrow
\psi^{\pm (1)}_{n}$ as $k \rightarrow \infty$
and $\psi^{\pm (1)}_{n} - \psi^{\pm (k)}_{n} \in K_{\pm}({\bf
L}^1)$.  This is somewhat short of our goal, but
illustrates the intermediate result that the kernel is not
closed.  We also see that ${\bf L}^1$ is not continuous
(even on ${\cal D}_{\pm {\bf L}^1}$) as a map from ${\cal D}_{\pm {\bf L}^1}$
with the ${\bf {\cal L}}^2$ topology to functions of integers
with the topology of pointwise convergence.  This
accounts for many of the strange features of the loop
transform.

We are now prepared to construct the function
$\psi_*^{\Lambda^{\pm}, (k)}$ that will be shown to take on the same
values as $\Lambda^{\pm}_*(s)$ at integer points for a continuous
function $\Lambda^{\pm}(a)$
of compact
support\footnote{We will choose $\Lambda^{\pm}
\in {\cal C}^0_{0\pm}({\cal R})$.}.
For any such
$\Lambda^{\pm} \in {\bf {\cal L}}^2_{\pm}({\cal R})$, consider the
function

\begin{equation}
\label{psisum}
\psi_*^{\pm \Lambda , (k)}(s) \equiv \sum^{\infty}_{m=0}
\hat{\Lambda}^{\pm} (m) \psi^{\pm (k)}_{*m}(s)
\end{equation}
defined for all $s \in {\bf {\cal C}}$ for which this sum
converges pointwise.  Our first task will be to show that
this sum actually converges pointwise for all
complex $s$.  Now, $\Lambda^{\pm}$ has compact support
that must lie in some $(-L, L) \subset {\bf {\cal R}}$.
By continuity and compactness of the support, $|\Lambda
^{\pm}|$ takes on some maximal value, $\Lambda
^{\pm}_{max}$.  Thus,

\begin{equation}
|\hat{\Lambda}^{\pm}(m)| \leq 2L \Lambda^{\pm}_{max}
e^{|m|L}
\end{equation}

If we use

\begin{equation}
\eqnum{fact iii}
\label{factiii}
\Bigg| {{\sin{(\pi z)}} \over {\pi z}} \Bigg| \leq
{{\sinh {(\pi |z|)}} \over {\pi |z|}} \leq e^{\pi |z|}
\end{equation}
for all complex, $z = s+it$, we find an estimate of the magnitude of
$\psi_{*n}^{\pm (k)}(s+it)$:

\begin{equation}
\label{bound}
|\psi_{*n}^{\pm (k)}(s+it)| \leq 2^k e^{-k \alpha n^2}
e^{k(\alpha s + 2 \pi |s|)} e^{k(- \alpha t^2 + 2 \pi
|t|)} e^{2 \pi kn}
\end{equation}
for all $s,t \in {\bf {\cal R}}$.  It follows that the
sum in \ref{psisum} converges for all complex $s$ and
does so uniformly in any compact region, $K \subset {\bf
{\cal C}}$.  In fact,

\begin{equation}
|\psi^{\Lambda_{\pm},(k)}_*(s)| \leq 2^{k+1} L
\Lambda^{\pm}_{max} \biggl[ \exp{\biggl({{(2 \pi k + L)^2} \over {2 \alpha
k}}\biggr)} + 1 \biggr] e^{k(\alpha s^2 + 2 \pi |s|)} e^{k(- \alpha t^2 + 2
\pi |t|)} \sqrt{\pi}
\end{equation}
so that $|\psi^{\Lambda^{\pm},(k)}_*(s)|$ is integrable
along every contour parallel to the imaginary axis and
satisfies the limit condition of the transform inversion
theorem.

In order to use the transform inversion theorem,
all that remains is to show that $\psi^{\Lambda^{\pm},(k)}_*(s)$ is
analytic on the entire complex plane.  We already know
that this function is defined for all complex $s$, so we
need only show that it satisfies the Cauchy-Riemann
equations. Since the sum in \ref{psisum} converges
uniformly in every compact subset of the complex plane,
we can take derivatives of $\psi^{\Lambda^{\pm},(k)}_*(s)$
term by term in the sum so long as the sum of the
derivatives converges for all complex $s$.  Use of our
three useful facts shows that, for $n>2s$.

\begin{equation}
\bigg| \case{d}{ds} \psi^{\pm (1)}_{*m} \bigg| < 2 \alpha
|s| \psi^{\pm (1)}_{*m} + 4 e^{- \alpha n^2} e^{\alpha s^2}
e^{2 \pi |s|} e^{2 \pi n} \Biggl({{4 e^{|s|}} \over {\pi}}
\Biggr)
\end{equation}
so that (using Eq. \ref{bound} as well) the sum converges
absolutely for all complex $s$.  Since each
$\psi^{\pm (k)}_{*m}(s)$ satisfies the Cauchy-Riemann
equations for all complex $s$, so does
$\psi^{\Lambda^{\pm} (k)}_*$, and $\psi^{\Lambda^{\pm} (k)}_*$
is a holomorphic function on ${\bf {\cal C}}$.  We see
that $\psi^{\Lambda^{\pm} (k)}_*$ satisfies all of the
conditions of the transform inversion theorem, so there
must exist some $\psi^{\Lambda^{\pm} (k)}$ such that
the two-sided Laplace transform of $\psi^{\Lambda^{\pm} (k)}$
is $\psi^{\Lambda^{\pm} (k)}_*$.
Using $||\tilde{\psi}^{\pm (k)}_{m}||_{{\bf {\cal L}}^2}$
to estimate $||\tilde{\psi}^{\Lambda^{\pm} (k)}||_{{\bf
{\cal L}}^2}$ and thus $||\psi^{\Lambda^{\pm} (k)}||_{{\bf
{\cal L}}^2}$, we see that $\psi^{\Lambda^{\pm} (k)} \in
{\bf {\cal L}}^2_{\pm}({\bf {\cal R}})$ (for odd k) and
that

\begin{equation}
||\psi^{\Lambda^{\pm} (k)}||^2_{{\bf {\cal L}}^2} \leq
{{8L^2 \Lambda^{\pm 2}_{max}} \over {\sqrt{k}}} \biggl(
\sqrt{{{\pi} \over {4 \alpha k}}} e^{L^2/2 \alpha k} +1 \biggr)
\end{equation}
which approaches zero as $k \rightarrow \infty$.

Finally, we define $\phi^{\Lambda^{\pm} (k)} \equiv
\Lambda^{\pm} - \psi^{\Lambda^{\pm} (k)}$, and find that
$\phi^{\Lambda^{\pm} (k)} \rightarrow \Lambda^{\pm}$ in
${\bf {\cal L}}^2_{\pm}({\cal R})$ as $k \rightarrow \infty$ and

\begin{eqnarray}
\hat{\phi}^{\Lambda^{\pm} (k)}(n) &=& \hat{\Lambda}^{\pm}(n)
- \hat{\psi}^{\Lambda^{\pm} (k)}(n) = \hat{\Lambda}^{\pm}(n)
- \psi^{\Lambda^{\pm} (k)}_*(n) \cr
&=& \hat{\Lambda}^{\pm}(n)
- \sum^{\infty}_{m=0} \hat{\Lambda}^{\pm (m)} \psi^{\pm (k)}_{*m}(n) \cr
&=& \hat{\Lambda}^{\pm}(n) - \sum^{\infty}_{m=0}
\hat{\Lambda}^{\pm (m)} \delta_{m,n} =
\hat{\Lambda}^{\pm}(n) - \hat{\Lambda}^{\pm}(n) = 0
\end{eqnarray}
so that $\phi^{\Lambda^{\pm},(k)} \in K_{\pm}({\bf L}^1)$
and we see that ${\bf {\cal L}}^2_{\pm}({\cal R}) \subset
\overline{{\cal C}^0_{0\pm}({\bf {\cal R}})} \subset
\overline{K_{\pm}({\bf {\cal L}}^1)}$ and $K_{\pm}({\bf L}^1)$ is dense in
${\bf {\cal L}}^2_{\pm}({\cal R})$.  It follows that, since
$K({\bf L}) = K_+({\bf L}^1) \otimes K_+({\bf L}^1) \ \oplus \
K_-({\bf L}^1) \otimes K_-({\bf L}^1)$, we have
\begin{eqnarray}
\overline{K({\bf L})} &=& \overline{K_+({\bf L}^1)} \otimes
\overline{K_+({\bf L}^1)} \ \oplus \
\overline{K_-({\bf L}^1)} \otimes \overline{K_-({\bf L}^1)} \cr
&=& {\cal L}^2_({\cal R}) \otimes {\cal L}_+^2({\cal R}) \ \oplus \
{\cal L}^2_-({\cal R}) \otimes {\cal L}^2_-({\cal R}) \cr
&=& {\cal L}^2_+ ({\cal R}^2)
\end{eqnarray}
and we have shown that the kernel of the loop transform is dense in the
connection representation.

Note that the only fact about ${\Lambda}^{\pm}$ that we actually used in
our construction of the function $\psi^{\Lambda^{\pm},(k)}(a)$ was that
$|\hat{\Lambda}^{pm}(m)| \leq 2L \Lambda_{max}^{\pm} e^{|m|L}$.  It follows
that given any set of numbers $\{ \hat{\psi}_1(n)\}$ that satisfy
$\hat{\psi}_1(n) = \pm \hat{\psi}_1(-n)$ and also have a bound of
the form $|\hat{\psi}_1(n)| \leq \psi_{max} e^{|n|L}$ for some $\psi_{max}$
and $L$, there is some $\psi_1 \in {\cal L}^2_{\pm}$ such that $({\bf L}^1
\psi_1)(n) = \hat{\psi}_1(n)$.  If instead of working with functions of
one dimensional arguments we work with functions of two dimensional
arguments:
\begin{equation}
\hat{\psi}(n,m) = \hat{\psi}(-n,-m) \text{ and} |\hat{\psi}(n,m)| \leq
\psi_{max} e^{|n|L_1 + |m| L_2}
\end{equation}
for some $\psi_{max}, L_1, L_2 \in {\cal R}$, we could define $\psi_*$
by
\begin{equation}
\psi_*(s) \equiv \sum^{\infty}_{n,m=0}
\hat{\psi}(n,m) \psi_{*n}(s^1) \psi_{*m}(s^2)
\end{equation}
by analogy with Eq. \ref{psisum} then a similar calculation would show
that there must be some function $\psi({\bf a})$ in ${\cal L}^2_+({\cal R})$
such that $\hat{\phi}(n,m)$ is the loop transform of $\psi({\bf a})$.
In particular, this shows that the image of the loop transform contains
all $l^2$ functions of loops; that is, all of those functions that make
up the standard loop representation.

\section{Some Candidate Spaces}
\label{Vs}

In this appendix we show that there do, in fact, exist
two subspaces that satisfy the requirements given in
section \ref{looprep} and give proofs of the properties
of these subspaces that were quoted.  That is, we seek linear subspaces
${\bf {\cal V}}$ of ${\bf {\cal L}}^2_+({\bf {\cal
R}}^2)$ that satisfy:

i) ${\bf {\cal V}} \rm{ \ is \ dense \ in \ } {\bf {\cal
L}}^2_+$

ii) ${\bf {\cal V}} \rm{ \ is \ stable \ under \ } {\cal T}^0,
{\cal T}^1\rm{'s}$

iii) ${\bf {\cal V}} \cap K({\bf L}) = \{0\}$

iv) ${\bf {\cal V}} \subset {\cal D}_{\bf L}$

For the first such subspace, let ${\bf {\cal V}}_1$
be the vector space consisting of all functions of the
form:

\begin{eqnarray}
\label{def1}
\psi_{{\bf A},{\bf B}} = e^{-|{\bf a}|^2} \sum_{{\bf m} \in
{\cal Z}^2/{\cal Z}_2, {\bf n} \in
{\cal Z}^+ \times {\cal Z}^+} \biggl(
&\cosh&{({\bf m} \cdot {\bf a})} (a_1)^{n_1}(a_2)^{n_2}
A_{{\bf m},{\bf n}} \cr + &\sinh&{({\bf m} \cdot {\bf a})}
(a_1)^{n_1}(a_2)^{n_2} B_{{\bf m},{\bf n}} \biggr)
\end{eqnarray}
where $A_{{\bf m},{\bf n}} = 0$ for $n_1
+ n_2$ odd and $B_{{\bf m},{\bf n}} = 0$ for $n_1 + n_2$
even.  The coefficients $A_{{\bf m},{\bf n}}$ and $B_{{\bf m},{\bf n}}$
are to be nonzero only for a finite number of pairs ${\bf n}$, ${\bf m}$.

To see that this space is
stable under the ${\cal T}^0$'s and ${\cal T}^1$'s,
note that taking derivatives and
multiplication by cosh and sinh preserves the form
described in Eq. \ref{def1} up to possible violations of the conditions on the
coefficients that depend on the even or odd character of $n_1+n_2$.
However, the ${\cal T}^0$'s and ${\cal T}^1$'s preserve ${\cal L}^2_+$, so
these conditions must not be violated.  The space of functions described
by Eq. \ref{def1} is therefore invariant under the action of the
${\cal T}^0$'s and ${\cal T}^1$'s.

We can also see that ${\bf {\cal V}}_1$ is dense in ${\bf {\cal L}}^2_+$ since
it contains the (properly symmetrized) energy eigenstates
of two uncoupled harmonic oscillators (i.e, the states of
the form $e^{-|{\bf a}|^2}H_{n_1}(a_1)H_{n_2}(a_2)$)
where $H_n$ is the Hermite polynomial of order $n$).
The rapid fall off of $e^{-|{\bf a}|^2}$ ensures that ${\bf {\cal
V}}_1 \subset {\cal D}_{\bf L}$.  Thus, we need only show
that ${\bf {\cal V}}_1$ satisfies (iii).

To verify this final requirement, we note that the two sided Laplace transform
of $\psi_{{\bf A},{\bf B}}$ is of the form,

\begin{eqnarray}
\label{decomp1}
\psi_{{\bf A},{\bf B}} = e^{-|{\bf s}|^2/4}
\sum_{\scriptstyle {\bf m} \in {\bf {\cal
Z}}^2/{\bf {\cal Z}}_2, {\bf n} \in {\cal Z}^+ \times {\cal Z}^+}
& \biggl( &  \cosh{\Bigl({{{\bf m} \cdot {\bf
s}} \over {2}}\Bigr)} {\cal P}_{n_1}^{(c)}(s_1){\cal
P}_{n_2}^{(c)}(s_2) A_{{\bf m},{\bf n}} \cr
&+& \sinh{\Bigl({{{\bf m} \cdot
{\bf s}} \over {2}}\Bigr)} {\cal P}_{n_1}^{(s)}(s_1){\cal
P}_{n_2}^{(s)}(s_2) B_{{\bf m},{\bf n}} \biggr)
\end{eqnarray}
 where ${\cal
P}_{n}^{(c)}(s)$, ${\cal
P}_{n}^{(s)}(s)$ are polynomials of degree $n$ in $s$.
Any function of this form with a finite number of nonzero
coefficients has a largest $m_1$ such that $A_{{\bf
m},{\bf n}}$ or $B_{{\bf m},{\bf n}}$ is nonzero for some
$m_2, n_1, n_2$ and a largest $n_1$ such that (for this
value of $m_1$ and some $m_2, n_2$) $A_{{\bf m},{\bf n}}$
or $B_{{\bf m},{\bf n}}$ is nonzero.  For fixed $s_2$ and
large enough $s_1$, this term will dominate and
$\psi_{{\bf A},{\bf B}}(s_1,s_2)$ is not zero unless

\begin{equation}
\label{depends}
\sum_{n_2 =0, m_2 = -\infty}^{\infty} e^{m_2s_2/2} \biggl(
{\cal P}_{n_2}{\cal
P}_{n}^{(c)}(s)(s_2) A_{{\bf
m},{\bf n}}C_1 + {\cal P}_{n_2}{\cal
P}_{n}^{(s)}(s)(s_2) B_{{\bf
m},{\bf n}} C_2 \biggr) = 0
\end{equation}
for some non-zero $C_1, C_2 \in {\bf {\cal R}}$.

Now, there must be at least one set of values
$(m_1,n_1)$ that are selected by this procedure for
values of $s_2$ that make up an unbounded subset $S$ of
${\bf {\cal R}}^+$.  Consider some $s_2$ in this set and
those values, $m_1$ and $n_1$.  We now take the largest
$m_2$ and $n_2$ that contribute in \ref{depends} and find
that, for large enough $s_2$ in our set $S$, this term
dominates and $\psi_{{\bf A},{\bf B}}(s_1,s_2)$ is not
zero unless $C_3 A_{{\bf m},{\bf n}} + C_4 B_{{\bf
m},{\bf n}} = 0$ for some non-zero $C_3,C_4 \in {\bf {\cal
R}}$.  This cannot happen since $A_{{\bf m},{\bf n}}$ and
$B_{{\bf m},{\bf n}}$ cannot both be non-zero for the same
choice of ${\bf n}$ by our symmetry assumption, though by
construction at least one of them is non-zero.  It follows that
${\cal V}_1$ satisfies all of the stated requirements.

The second example of a subspace satisfying (i)-(iv) is
quite similar.  Basically, it consists of the same set of
functions multiplied by judicious choices of sine and
cosine functions. We consider the set of functions that
can be obtained from
\begin{equation}
\label{v2start}
\psi_{\alpha} = \Re{[s^{-\alpha (a_1 + \case{i}{
\alpha})^2}]} \ \Re{[s^{-\alpha (a_2 + \case{i}{
\alpha})^2}]}
\end{equation}
for $\alpha = m/4 \pi$, $m \in {\bf {\cal Z}}^+$ by
multiplication with polynomials that are even under ${\bf
a} \rightarrow - {\bf a}$ and the functions that can in
turn be obtained from these by acting with any finite
number of ${\cal T}^0$'s and ${\cal T}^1$'s.  Let ${\bf {\cal
V}}_2$ be the set of finite linear combinations of these
functions.  By inspection, we see that any such function
can be written in the form
\begin{eqnarray} \psi ({\bf a})
= \sum_{\scriptstyle {i,k \in {\bf {\cal Z}}^+ \cup \{0\}}
\atop \scriptstyle {m \in {\bf {\cal Z}}^+, \ j,n \in
{\bf {\cal Z}}}} C_{i,j,k,m,n,} &\Re&{[(a_1)^ie^{ja_1}
\exp{(-\case{m}{4 \pi} (a_1 + \case{2i \pi}{m})^2)}]} \cr
\times &\Re&{[(a_2)^ke^{na_2} \exp{(-\case{m}{4 \pi} (a_2 +
\case{2i \pi}{m})^2)}]}
\end{eqnarray}

Since taking the Laplace transform is a real operation,
the (two-sided) Laplace transforms of these functions are
of the form

\begin{eqnarray}
\label{depends2}
\psi ({\bf a}) =  \sum_{\scriptstyle {i,k \in {\bf {\cal Z}}^+ \cup \{0\}}
\atop \scriptstyle {m \in {\bf {\cal Z}}^+, \ j,n \in
{\bf {\cal Z}}}} C_{i,j,k,m,n,}
&\Re&{[{\cal P}_i(s_1)e^{js_1/2} \exp{(\case{m}{16 \pi}
(s_1 + \case{4i \pi}{m})^2)}]} \cr
\times &\Re&{[{\cal
P}_k(s_2)e^{ns_2/2} \exp{(\case{m}{16 \pi} (s_2 + \case{4i
\pi}{m})^2)}]}
\end{eqnarray}
where again ${\cal P}_n(s)$ is some polynomial of degree
$n$ in $s$ (with a real leading coefficient).  The term with the smallest $m$
value in \ref{depends2} must decay the most slowly
and eventually dominate.  An imaginary change of origin
does not effect the argument we used to show that none of
the nontrivial elements of ${\bf {\cal V}}_1$ are in
$K({\bf L})$, so we may use it again here on the term
with least $m$ to show that no nontrivial element of
${\bf {\cal V}}_2$ is in $K({\bf L})$.

We still need to show that ${\bf {\cal V}}_2$ is dense in
${\bf {\cal L}}^2_+$, but we know that we can
generate all functions of the form:

\begin{equation}
\Re[{\cal P}_{n_1}(a_1 + \case{4 \pi i}{ m}) \exp{(-
\case{m}{4\pi} (a_1 + \case{4 \pi i}{m})^2)}] \ \Re[{\cal
P}_{n_2}(a_2 + \case{4 \pi i}{m}) \exp{(- \case{m}{4\pi}
(a_2 + \case{4 \pi i}{m})^2)}] \end{equation}
where ${\cal P}_n$ is an even polynomial of degree n.
Using the convergence of the power series expansion for
$e^{x^2}$, the rapid decay of $e^{-x^2}$,
and the fact that $\Re[\exp{(- \case{m}{4\pi} (a + \case{4
\pi i}{m})^2)}]$ is even in $a$, we can conclude that

\begin{eqnarray}
&\Re&[\exp{(k_1 (a_1 + \case{4 \pi i}
{m})^2)} \exp{(- \case{m}{4\pi} (a_1 + \case{4 \pi
i}{m})^2)}] \cr \times &\Re&[\exp{(k_2 (a_2 + \case{4 \pi i}{m})^2)}
\exp{(- \case{m}{4\pi} (a_2 + \case{4 \pi i}{m})^2)}] \in
\overline{{\bf {\cal V}}}_2 \end{eqnarray} for any $k_1,
k_2 < \case{m}{4\pi}$.  In particular, let $k_1,k_2 =
\case{m}{4\pi} - 1$.  Then we have shown that $\Re[\exp{(-
(a_1 + \case{4 \pi i}{m})^2)}] \ \Re[\exp{(-
(a_2 + \case{4 \pi i}{m})^2)}] \in \overline{{\bf {\cal
V}}}_2$.  Finally, in this form we can take the limit
$m \rightarrow \infty$ so that we find that $e^{-
(a_1)^2} e^{- (a_2)^2} \in \overline{{\bf {\cal V}}}_2$.
We could similarly obtain this function multiplied by any
polynomial that is even in ${\bf a}$.  The set of such
functions is a dense subspace of ${\bf {\cal
L}}^2_+$ which shows that ${\bf {\cal V}}_2$ is dense in
${\bf {\cal L}}^2_+$.

Thus we see that ${\bf {\cal V}}_2$ also satisfies
requirements (i)-(iv).  Note that it has no
elements in common with ${\bf {\cal V}}_1$.  However,
their loop transforms do have elements in common.  To show
this, we consider the Laplace transform of one of the
functions,  $\psi_{\case{1}{4
\pi}}$, defined in \ref{v2start}:

\begin{equation}
\psi_{*\case {1}{4 \pi}} = C \Re [e^{(s_1 + 4i \pi)^2/4}] \
\Re [e^{(s_2 + 4i \pi)^2/4}] = C e^{- 4 \pi}
e^{((s_1)^2+(s_2)^2)/4} \cos{(2 \pi s_1)} \cos{(2 \pi
s_2)}
\end{equation}
for some $C \in {\bf {\cal R}}$.  For
integer $s$, the cosines are just equal to one.  Up to
a constant factor, this is the same as the two-sided
Laplace transform of $e^{- (a_1)^2} e^{- (a_2)^2}$, which
is in ${\bf {\cal V}}_1$. The images of ${\bf {\cal V}}_1$
and ${\bf {\cal V}}_2$ have elements in common, as was
claimed in section \ref{looprep}.  A careful calculation shows
that the functions in ${\cal V}_1$ and ${\cal V}_2$ that map to the
same function of loops have different ${\bf {\cal L}}^2$ norms.

It was also mentioned in section \ref{looprep} that neither
of these spaces have subspaces that fulfill
requirements (i)-(iv) and which form an algebraically
irreducible representation of the ${\cal T}^0$'s and ${\cal
T}^1$'s.  This is because acting with
a ${\cal T}^1({\bf k})$ increases the degree of the
polynomial part of the function according to the
decompositions used in Eq. \ref{decomp1} and Eq. \ref{depends2} but no
action by any combination of ${\cal T}^0$'s and ${\cal T}^1$'s
can lower the polynomial degree of these functions.  This is the
last of the properties of these subspaces that was quoted in
section \ref{looprep}.

\section{The Inner Product in the Connection Representation}
\label{Tip}

This appendix serves to verify that the
${\cal T}^0$'s and ${\cal T}^1$'s determine the same inner product
in the connection representation as ${\bf a}$ and ${\bf p}$.
With the assumption that
the connection representation forms an ${\bf {\cal L}}^2$
space, we will show that Hermiticity of
the ${\cal T}^0$'s and ${\cal T}^1$'s requires the same inner product as
Hermiticity of ${\bf a}$ and ${\bf p}$ if the representation is
required to be topologically irreducible.  Interestingly, topological
irreducibility plays a much stronger role when the inner product is
determined using loop variables than when ${\bf a}$ and ${\bf p}$
are used.

In the connection representation, ${\cal T}^0({\bf k})$
is represented by a multiplication operator and ${\cal
T}^1({\bf k})$ is represented by a differential
operator.  Thus, in the timelike sector,
we consider an inner product defined on
some set of smooth functions on the torus.  Since the
configuration space is a compact manifold without boundary, we will be able
to perform integrations by parts in the integral that
defines the inner product.  Performing these integrations
by parts shows that we must have

\begin{equation}
\sin{({\bf n} \cdot {\bf a})} \ \ {\bf n} \times
\case{\partial}{\partial {\bf a}} d\mu = 0
\end{equation}
where $d \mu$ is the measure that defines the inner
product, for the ${\cal T}^1({\bf n})$'s to be
Hermitian.  This requires the measure
to be constant, except at the points $a_b \in \{0, \pi\}$
where it could have delta-function
singularities\footnote{Once again, thanks to Jorma
Louko for pointing this out.}.
If it does have delta function singularities at these
points, then the completion of this space is ${\bf {\cal
L}}^2 \oplus {\bf {\cal C}}^4$, where the ${\bf {\cal
C}}^4$ is represented by characteristic functions of the four
singular points.  These functions are mapped into
themselves by the ${\cal T}^0$'s and ${\cal T}^1$'s, and so
form a closed invariant subspace and the representation
is not topologically irreducible.  The only choice that
gives a topologically irreducible representation is
to include none of these singularities in $d \mu$ so that
$d \mu$ must be a constant.

Note that we could also build a topologically irreducible representation
by starting with the set of all smooth functions on the torus that
vanish at these four exceptional points.  There is then no ambiguity
in the measure; it must be constant.  However, the
set of functions that vanish at a discrete set of point is dense
in ${\cal L}^2$, so the completed space is the same as the one given
by the above argument, ${\cal L}^2({\cal T}^2)$.

The same argument can be performed
in the spacelike sector by replacing $\sin{({\bf n} \cdot {\bf a})}$
with $\sinh{({\bf n} \cdot {\bf a})}$ in the above discussion and
by considering an initial set functions on the configuration space
that contains
all smooth functions that fall off
sufficiently rapidly near infinity.  In this case, only the origin
appears as an exceptional point so that there is only one free
parameter in the measure corresponding to the strength of a
delta-function singularity at the origin.  If this parameter is
not set to zero, the completion of our initial set of functions yields
${\cal L}^2 \oplus {\cal C}$, which is again topologically reducible.
The only topologically irreducible choice is to set this parameter to
zero and to find that the completed Hilbert space is just ${\cal L}^2$.
Again, we would uniquely arrive at this space without
explicitly imposing topological irreducibility if we started with
a space of smooth functions that vanish at the exceptional point.

\section{The Uniqueness of the Loop Representation}
\label{uip}

In this appendix we will show that there is a unique definition of the
loop representation and a unique
choice of inner product for this representation such that

\par

i) The loop representation contains at least one function with
support on a single loop.

ii) The function of loops\footnote{In this appendix
the word ``loop" will again be used to refer to equitopy classes
of loops.} that maps all loops to zero is the zero
state.

iii) The ${\cal T}^0$'s and ${\cal T}^1$'s are Hermitian with
respect to the inner product.

iv) The loop representation is a topologically irreducible
representation of the ${\cal T}^0$'s and ${\cal T}^1$'s.

\par

We will begin by showing that our loop representation ${\cal R}$
contains all functions with support on a finite number of loops.  Note
that we are guaranteed by (i) to have at least one such function that
in fact has support on only one loop, ${\bf n}$.  Let's denote by
$|{\bf n}\rangle$ the
function that takes the value one on this loop and zero on all
others.  We similarly define $|{\bf k}\rangle$ for all other loops, ${\bf k}$,
although at this point we do not know if ${\cal R}$ contains any of these
other functions.

To show that these other functions with support on single loops are in
${\cal R}$, we consider the action of the
operator $\case{1}{2} ({\cal T}^1({\bf k}) + i ({\bf n} \times {\bf k} \
{\cal T}^0
({\bf k}))$ on $|{\bf n}\rangle$:
\begin{equation}
\case{1}{2} ({\cal T}^1({\bf k}) + i ({\bf n} \times {\bf k} \ {\cal T}^0
({\bf k}))|{\bf n}\rangle = i({\bf n} \times {\bf k}) \ |{\bf n} + {\bf k}
\rangle
\end{equation}
Since $R$ is a representation of the ${\cal T}^0$'s and ${\cal T}^1$'s, this
implies that for ${\bf k} \times {\bf n} \neq 0$ $R$ contains
$|{\bf n} + {\bf k} \rangle$.  However,  ${\bf k} \times {\bf n} \neq 0$
is equivalent to  $({\bf k} - {\bf n}) \times {\bf n} \neq 0$ so that
for  ${\bf k} \times {\bf n} \neq 0$ $R$ contains
$|{\bf k} \rangle$.  If we choose some such ${\bf k}$, then $R$ must
also include the state $|2{\bf k}\rangle$ and therefore the state:
\begin{equation}
{\cal T}^0({\bf k})|{\bf k}\rangle - \case{1}{2} |2 {\bf k} \rangle =
\case{1}{2} |{\bf 0}\rangle
\end{equation}
so that the characteristic function of the zero loop is a member of our
representation.  Since ${\cal T}^0({\bf k})|{\bf 0}\rangle = |{\bf k}
\rangle$, all characteristic functions of loops and therefore all
functions of loops with finite support are elements of our loop
representation.

To show the uniqueness of the inner product in the loop
representation, we will need a few simple relations that
follow from the definitions of the ${\cal T}^0$'s and ${\cal
T}^1$'s.  We will denote the inner product of
$|{\bf n}\rangle$ and $|{\bf m}\rangle$ by
$\langle {\bf n}| {\bf m} \rangle$.  It follows that,
if the ${\cal T}^0({\bf k})$'s are hermitian:

\begin{equation}
\label{0I}
\langle {\bf m}| {\bf n} \rangle = \langle {\bf 0}|{\cal
T}^0({\bf m})| {\bf n} \rangle = \case {1}{2} [\langle
{\bf 0} | {\bf n} + {\bf m} \rangle + \langle {\bf 0} |
{\bf n} - {\bf m} \rangle] \end{equation}
so that all inner products of states which are given
by functions of loops are determined by the inner
products of the states $|{\bf n}\rangle$
with the state $|{\bf 0}\rangle$.

Also note that ${\cal T}^1({\bf k})$ annihilates $|{\bf
0}\rangle$ so that if ${\cal T}^1({\bf k})$ is Hermitian
\begin{equation}
\label{I}
0 = \langle {\bf 0}|{\cal T}^1({\bf n})|{\bf m}\rangle =
i{\bf n} \times {\bf m }(\langle {\bf 0}|{\bf n} + {\bf
m} \rangle - \langle {\bf 0}| {\bf n} - {\bf m} \rangle)
\end{equation}
and that if ${\bf n}$ and ${\bf m}$ are not parallel
as vectors in ${\cal Z}^2$ we have

\begin{equation}
\label{II}
\langle {\bf 0} | {\bf n} + {\bf m} \rangle = \langle
{\bf 0} | {\bf m} - {\bf n} \rangle
\end{equation}
We can use this equation to calculate all inner products of
the form $\langle {\bf n} |{\bf 0}\rangle$.  To do so, note that
any ${\bf n} \neq 0$ can be written in the form ${\bf n} =
{\bf k} + (a,b)$ where ${\bf k} \in 2{\bf {\cal Z}} \times
2 {\bf {\cal Z}}$ and $(a,b) \in
\{(0,1),(1,0),(1,1),(0,2)\}$.  If we consider some ${\bf n}$
that is not proportional to its corresponding $(a,b)$ then,
from Eq. \ref{II} we find that

\begin{equation}
\langle {\bf 0} | {\bf n} \rangle = \langle {\bf 0} |
{{{\bf n} + (a,b)} \over {2}}  + {{{\bf n} - (a,b)} \over
{2}} \rangle = \langle {\bf 0} |
{{{\bf n} + (a,b)} \over {2}}  - {{{\bf n} - (a,b)} \over
{2}} \rangle = \langle {\bf 0}| (a,b) \rangle
\end{equation}
If, on the other hand, ${\bf n}$ is proportional to its
$(a,b)$ we can perform this argument replacing $(a,b)$ with
$(a',b') = (a,b) + (4,2)$ to which ${\bf n}$ will not
then be proportional.  In this case, $\langle {\bf 0} | {\bf n} \rangle
= \langle {\bf 0}|
(a',b') \rangle$.  But, by the above argument,
$\langle {\bf 0} | (a,b) \rangle = \langle {\bf 0}|
(a',b') \rangle$.  If we fix the normalization of the zero
loop to be $\langle {\bf 0}| {\bf 0}\rangle = 1$, we
find that the inner products are determined by the four
parameters $\langle {\bf 0}|(a,b)\rangle$.

Under the
Fourier transform map from sequences to functions on the
torus, these four coefficients are just combinations of
the strengths of the four singularities allowed in the
measure, $d\mu$, of the connection representation
of the timelike sector that were found in
Appendix \ref{Tip}.  In
terms of loop coefficients, these parameters produce
terms in the inner product $\langle \psi_1|\psi_2 \rangle$
that depend on $\sum_{\scriptstyle {\bf n} \in 2{\bf {\cal
Z}} \times 2{\bf {\cal Z}} +(a,b)} \psi_1({\bf n})$
and  $\sum_{\scriptstyle {\bf n} \in 2{\bf {\cal
Z}} \times 2{\bf {\cal Z}} +(a,b)} \psi_2({\bf m})$.  The
subspace such that this sum is zero for all $(a,b)$ is
invariant under the action of the ${\cal T}^0$'s and ${\cal
T}^1$'s (in fact, it contains the image of every ${\cal
T}^1({\bf k})$).  When this term enters into the inner
product with non-zero coefficient, this subspace is closed
and the representation is not topologically irreducible\footnote{Thanks
to Alan Rendall for pointing this out.}.
Thus, if the representation is to be topologically
irreducible, these four parameters must be zero and the
inner product is unique.  This is just the result that would be
anticipated from Ref. \cite{Rendall}.

Since the inner products of the states $|{\bf n}\rangle$ are unique, $R$
must contain the completion in this inner product
of the space of functions with support
on a finite number of loops.  With the parameters $\langle {\bf 0} |
(a,b) \rangle$ set to zero, we see from Eq. \ref{0I} that $\langle {\bf n} |
{\bf m} \rangle = \case{1}{2} \delta_{{\bf n}, {\bf m}}$ for ${\bf n}$
or ${\bf m}$ not zero and $\langle {\bf 0}|{\bf 0}\rangle = 1$.
Therefore, this completion is just the set of $l^2$ functions of
loops.  However, this space is a closed subspace (in this topology)
of $R$ and is left invariant by the ${\cal T}^0$'s and ${\cal T}^1$'s so
that, if $R$ is indeed topologically irreducible, this space must be all of
$R$.  It follows that the choice of $R$ and the inner product imposed
on it follow uniquely from requirements (i)-(iv).

\end{document}